\documentclass{article}

\usepackage{arxiv}

\usepackage[utf8]{inputenc} % allow utf-8 input
\usepackage[T1]{fontenc}    % use 8-bit T1 fonts
\usepackage{hyperref}       % hyperlinks
\usepackage{url}            % simple URL typesetting
\usepackage{booktabs}       % professional-quality tables
\usepackage{multirow}
\usepackage[table,xcdraw]{xcolor}
\usepackage[
singlelinecheck=false % <-- important
]{caption}
\usepackage{amsfonts}       % blackboard math symbols
\usepackage{nicefrac}       % compact symbols for 1/2, etc.
\usepackage{microtype}      % microtypography
\usepackage{lipsum}		% Can be removed after putting your text content
\usepackage{graphicx}
\usepackage{natbib}
\usepackage{doi}
\usepackage{bm}
\usepackage[varvw]{newtxmath}       % selects Times Roman as basic font
\usepackage{amsmath}
\mathchardef\mhyphen="2D % Define a "math hyphen"
\usepackage{booktabs}
\usepackage{multirow}
\usepackage{comment}

\usepackage{listings}
\usepackage{xcolor}

\definecolor{codegreen}{rgb}{0,0.6,0}
\definecolor{codegray}{rgb}{0.5,0.5,0.5}
\definecolor{codepurple}{rgb}{0.58,0,0.82}
\definecolor{backcolour}{rgb}{0.95,0.95,0.92}

\lstdefinestyle{mystyle}{
    backgroundcolor=\color{backcolour},   
    commentstyle=\color{codegreen},
    keywordstyle=\color{magenta},
    numberstyle=\tiny\color{codegray},
    stringstyle=\color{codepurple},
    basicstyle=\ttfamily\footnotesize,
    breakatwhitespace=false,         
    breaklines=true,                 
    captionpos=b,                    
    keepspaces=true,                 
    numbers=left,                    
    numbersep=5pt,                  
    showspaces=false,                
    showstringspaces=false,
    showtabs=false,                  
    tabsize=2
}

\title{Digital Twin-Centered Hybrid Data-Driven Multi-Stage Deep Learning Framework for Enhanced Nuclear Reactor Power Prediction}

\author{ {James ~Daniell}\\
	Nuclear Engineering and Radiation Science\\
	Missouri University of Science and Technology\\
	Rolla, MO 65409, USA \\
\And
	{Kazuma ~Kobayashi} \\
	Grainger College of Engineering\\ Nuclear, Plasma, \& Radiological Engineering\\ University of Illinois Urbana-Champaign\\
	Urbana, IL, USA \\
\And
 	{Ayodeji ~Alajo} \\
	Nuclear Engineering and Radiation Science\\
	Missouri University of Science and Technology\\
	Rolla, MO 65409, USA \\
\And
 	{Syed Bahauddin ~Alam} \\
	Grainger College of Engineering\\ Nuclear, Plasma, \& Radiological Engineering\\ University of Illinois Urbana-Champaign\\
	National Center for Supercomputing Applications\\
    Urbana, IL, USA\\
}

\renewcommand{\shorttitle}{\textit{"Digital Twin-Centered Multi-Stage Deep Learning Framework"} }

\hypersetup{
pdftitle={A template for the arxiv style},
pdfsubject={q-bio.NC, q-bio.QM},
pdfauthor={David S.~Hippocampus, Elias D.~Striatum},
pdfkeywords={First keyword, Second keyword, More},
}

\begin{document}
\maketitle

\begin{abstract}
The accurate and efficient modeling of nuclear reactor transients is crucial for ensuring safe and optimal reactor operation. Traditional physics-based models, while valuable, can be computationally intensive and may not fully capture the complexities of real-world reactor behavior. This paper introduces a novel hybrid digital twin-focused multi-stage deep learning framework that addresses these limitations, offering a faster and more robust solution for predicting the final steady-state power of reactor transients. By leveraging a combination of feed-forward neural networks with both classification and regression stages, and training on a unique dataset that integrates real-world measurements of reactor power and controls state from the Missouri University of Science and Technology Reactor (MSTR) with noise-enhanced simulated data, our approach achieves remarkable accuracy (96\% classification, 2.3\% MAPE). The incorporation of simulated data with noise significantly improves the model's generalization capabilities, mitigating the risk of overfitting. Designed as a digital twin supporting system, this framework integrates real-time, synchronized predictions of reactor state transitions, enabling dynamic operational monitoring and optimization. This innovative solution not only enables rapid and precise prediction of reactor behavior but also has the potential to revolutionize nuclear reactor operations, facilitating enhanced safety protocols, optimized performance, and streamlined decision-making processes. By aligning data-driven insights with the principles of digital twins, this work lays the groundwork for adaptable and scalable solutions in nuclear system management.

%Six classifier models are developed and tested, and compared for the first stage of the model, with an output representing a large band power output resolution. 

%Physical information is collected from the initial and final steady states of the transient to create a single trial. A dataset is generated using the generalized prompt jump equation to expand upon the data collected from the research reactor. Noise is added to the simulated data to reduce the likelihood of memorization. 

\end{abstract}

\section{Introduction}
Modeling is often used to monitor physical changes in a system or to determine the limitations of a design before it is implemented. Modeling software such as SCALE or TRACE has been specifically developed to model nuclear systems reliably without having to produce the system physically and allow for data collection at locations where sensors could not be feasibly placed in a real system. While these types of computational models have a high degree of accuracy, they are difficult to develop, tend to be system-specific, and are computationally intensive \cite{samal2020characterization}. More specifically, without using significant computational resources, these models have difficulty making predictions with the desired degree of accuracy in the amounts of time required to make some operational decisions. These types of models, while they have their drawbacks, are still indispensable to nuclear engineers and designers due to their physics-basing and high-fidelity nature.

Due to recent developments in computing techniques and resources, new types of models have been utilized across all engineering disciplines. Deep Neural Networks (DNNs) are machine learning models that have received significant attention in recent years due to their versatility and well-documented development \cite{massaoudi2021convergence}. DNNs have a variety of different internal architectures, which have independent strengths and weaknesses and are applied in different instances depending on the problem being addressed \cite{demuth2014neural}. These types of models were designed using brains as a model, in which each neuron is connected to another with a certain weight which can then be adjusted to determine the strength of the connection. The result is a data-driven approach in which the DNN fits an input to an output by feeding the information through the neurons, which are collected in a series of layers \cite{demuth2014neural}. DNNs are beneficial compared to other types of models due to their relatively low consumption of computing resources after training and their ability to predict complicated phenomena without having to model each specific physical aspect in the system \cite{lu2021deep}.

While DNNs are practical, they often avoid modeling the physics of the desired problem. These models typically function as direct input-to-output mapping versus true physics-based or probabilistic analysis \cite{leva2017analysis}. That is, DNNs are more analogous to empirical models or equations with the input variables functioning as an initial condition. Physics-based models derive equations from known phenomena, whereas DNN models approximate an output value based on what it has learned from examining previous data points, usually by fitting to a distribution \cite{bin2013application}. To function as a true digital twin \cite{yadav2023state,liu2024development,kobayashi2024deep,kobayashi2024explainable,kobayashi2024ai}, a model must integrate the strengths of data-driven methods with physics-informed insights, enabling it to not only predict system behavior but also adapt dynamically to operational changes in real-time. Furthermore, in many cases, DNNs are trained using data collected or generated from physics-based models \cite{kim2014prediction}. Furthermore, previous knowledge can be leveraged to produce models through processes such as transfer learning or multi-fidelity modeling, which can infuse known physics information into machine learning models \cite{pan2023transfer, chen2021learning, kabir2024transfer}.

Similar to other types of nuclear engineering models, the DNNs are numerical models usually developed for a singular purpose. For a given DNN model, an input and output are defined before the model is developed. This generally results in a model that performs one specific task \cite{sola1997importance}. In the nuclear industry, the primary utility of DNN models is for safety-based predictions or thermal-hydraulic system analysis \cite{bin2013application}. These areas of study are expected due to the design basis of nuclear reactors focusing on Loss of Coolant Accidents (LOCA) \cite{kim2014prediction,do2018prediction}. Additionally, the thermal-hydraulic system in a nuclear power plant is vital to both the safety and performance of the reactor. DNNs serve as excellent tools for determining the current state of flows or components in a thermal-hydraulic system, which might be difficult to examine otherwise \cite{radaideh2020neural}. When adapted to digital twin frameworks and systems \cite{yadav2023state}, these models can provide real-time insights into reactor state, enabling predictive diagnostics, anomaly detection, and operational optimization.

Predictive models are useful in nuclear systems for both operation and development, as demonstrated in our previous works \cite{kobayashi2024deep,kobayashi2024explainable,kobayashi2024ai}. DNNs serve as a valid option for each of these purposes for separate reasons. As the foundation of a digital twin \cite{yadav2023state,liu2024development}, DNNs can act as a real-time decision support system, offering operators immediate insights into reactor states and predictive analysis during incidents. By leveraging their minimal computational resource requirements post-training, these models can make fast predictions within the desired degree of accuracy \cite{el2021artificial, griesemer2023accelerating}. This would allow operators to obtain additional information during an incident or to predict the system state ahead of time, each of which would allow for quicker reactions and a more detailed knowledge of the system at the desired time. Predictive model applications often have safety concepts implicitly built in as well. While not specifically aimed to address a safety problem, additional information, efficient automation procedure, and improved operator interfacing make systems and procedures safer \cite{corrado2021human}. Combining these features into a synchronized framework, a digital twin offers a transformational leap in monitoring, control, and decision-making of nuclear reactor systems.
Predictive models are useful in nuclear systems for both operation and development, as demonstrated in previous works \cite{kobayashi2024deep,kobayashi2024explainable,kobayashi2024ai}. DNNs serve as a valid option for each of these purposes for separate reasons. As the foundation of a digital twin \cite{yadav2023state,liu2024development}, DNNs can act as a real-time decision support system, offering operators immediate insights into reactor states and predictive analysis during incidents. By leveraging their minimal computational resource requirements post-training, these models can make fast predictions within the desired degree of accuracy \cite{el2021artificial, griesemer2023accelerating}. This would allow operators to obtain additional information during an incident or to predict the system state ahead of time, each of which would allow for quicker reactions and a more detailed knowledge of the system at the desired time. Predictive model applications often have safety concepts implicitly built in as well. While not specifically aimed to address a safety problem, additional information, efficient automation procedure, and improved operator interfacing make systems and procedures safer \cite{corrado2021human}. By combining these features into a synchronized framework, a digital twin offers a transformational leap in nuclear reactor monitoring and control.
Predictive models are useful in nuclear systems for both operation and development, as demonstrated in our previous works \cite{kobayashi2024deep,kobayashi2024explainable,kobayashi2024ai}. DNNs serve as a valid option for each of these purposes for separate reasons. As the foundation of a digital twin \cite{yadav2023state,liu2024development}, DNNs can act as a real-time decision support system, offering operators immediate insights into reactor states and predictive analysis during incidents. By leveraging their minimal computational resource requirements post-training, these models can make fast predictions within the desired degree of accuracy \cite{el2021artificial, griesemer2023accelerating}. This would allow operators to obtain additional information during an incident or to predict the system state ahead of time, each of which would allow for quicker reactions and a more detailed knowledge of the system at the desired time. Predictive model applications often have safety concepts implicitly built in as well. While not specifically aimed to address a safety problem, additional information, efficient automation procedure, and improved operator interfacing make systems and procedures safer \cite{corrado2021human}. By combining these features into a synchronized framework, a digital twin offers a transformational leap in nuclear reactor monitoring and control.
Predictive models are useful in nuclear systems for both operation and development, as demonstrated in our previous works \cite{kobayashi2024deep,kobayashi2024explainable,kobayashi2024ai}. DNNs serve as a valid option for each of these purposes for separate reasons. As the foundation of a digital twin \cite{yadav2023state,liu2024development}, DNNs can act as a real-time decision support system, offering operators immediate insights into reactor states and predictive analysis during incidents. By leveraging their minimal computational resource requirements post-training, these models can make fast predictions within the desired degree of accuracy \cite{el2021artificial, griesemer2023accelerating}. This would allow operators to obtain additional information during an incident or to predict the system state ahead of time, each of which would allow for quicker reactions and a more detailed knowledge of the system at the desired time. Predictive model applications often have safety concepts implicitly built in as well. While not specifically aimed to address a safety problem, additional information, efficient automation procedure, and improved operator interfacing make systems and procedures safer \cite{corrado2021human}. By combining these features into a synchronized framework, a digital twin offers a transformational leap in nuclear reactor monitoring and control.
DNNs could also be useful for developing a full model by compiling multiple physical phenomena into a single input-output map. With the multiple different processes in a nuclear system affecting the system as a whole, the interactions between these processes can be difficult to model via physics-based equations \cite{kim1993application, pickering2022discovering}. Such an approach aligns with the principles of digital twins \cite{yadav2023state} by enabling real-time, system-wide predictions while remaining adaptable to changing conditions. DNNs could allow for each of these phenomena to contribute to the output and model the effects of their interactions numerically without having to develop a complicated physical model \cite{myung1991thermal,zhang2019thermal, pickering2022discovering}. In some disciplines, significantly robust models can be used to create a digital twin of a system \cite{yadav2023state}. A DNN could be feasibly used to remedy this by simplifying a complicated multi-physics approach when used in conjunction with other models \cite{shouman2022hybrid, chen2021learning}.

However, there are \textit{\textbf{challenges}} of DNNs for nuclear reactor modeling. \textbf{Research Gaps This Study is Filling}:
\begin{itemize}
    \item \textit{Overcoming Limitations of Existing Models: }The study addresses the limitations of both traditional physics-based models and conventional DNNs by proposing a multi-stage DNN framework. This approach combines classification and regression stages to improve prediction accuracy and robustness.
    \item \textit{Incorporation of Simulated Data with Noise:} To tackle the common issue of overfitting in nuclear system modeling, this study introduces the use of simulated data with noise. This method enhances the model's robustness and its ability to generalization \cite{kobayashi2024improved} across different scenarios, a significant gap in existing modeling approaches.
    \item \textit{Utilizing Real-World and Simulated Data: }The integration of real-world reactor operation data from the Missouri University of Science and Technology Reactor (MSTR) with simulated data is a novel approach that enhances the model's learning process and its applicability to real-world scenarios.
\end{itemize}

The objectives of the study are focused on enhancing the prediction of nuclear reactor behaviors through the development and implementation of a novel deep learning approach. Specifically, the paper aims to:

\begin{itemize}
    \item \textit{Develop a Novel Multi-Stage Deep Learning (DL) Framework:} Introduce a multi-stage DL framework for predicting the final steady-state power of reactor transients, leveraging both classification and regression stages within feed-forward neural networks. This framework is designed to improve prediction accuracy and efficiency over traditional and existing deep learning models.
 
    \item \textit{Combine Real-World and Simulated Data:} Utilize a unique dataset that combines real-world operational data from the Missouri University of Science and Technology Reactor (MSTR) with simulated, noise-enhanced data. This approach aims to tackle common issues such as overfitting and to enhance the model's robustness and ability to reflect real-world reactor behaviors accurately.

    \item \textit{Enhance Model Robustness Against Overfitting:} Address the challenge of overfitting—a common issue in nuclear system modeling—by introducing simulated data with noise into the training process. This method is intended to make the deep learning model more generalizable and reliable across different reactor operating scenarios.

    \item \textit{Predict Reactor Power Output with High Accuracy:} Ultimately, the objective is to predict reactor behavior efficiently, specifically the ending steady-state power output following reactor transients, with high levels of accuracy. The study seeks to demonstrate that the proposed multi-stage DL framework can achieve superior performance in this regard, offering a promising alternative for nuclear engineering applications.
 
    \item \textit{Improve Nuclear Reactor Operation and Safety: }By providing a tool for accurate and efficient prediction of reactor power states, the study aims to contribute to the broader goals of improving the operational efficiency and safety of nuclear reactors. This is achieved by enabling reactor operators to make faster, more informed decisions based on the model's predictions.
\end{itemize}

These objectives collectively target the advancement of nuclear reactor modeling and operational safety through the application of cutting-edge deep learning techniques. By addressing existing limitations and incorporating real-time predictive and prescriptive capabilities, this work lays the foundation for the development of a machine learning framework \cite{yadav2023state} for digital twin-based reactor operations, opening new avenues for research and application in the field.

In detail, the multi-stage modeling approach is developed to improve on the standard feed-forward design and leverage existing advancements in machine learning, with a clear alignment to digital twin principles. This style of model is developed by producing a classification result from a front-end model, which is then provided along with the input data to produce a regression prediction in the back-end model. This avoids complications associated with multi-task learning while still leveraging the benefits.  Such a structure is critical for ensuring the modularity, real-time inference, and adaptability expected from a digital twin. This structure is chosen due to some of the prevalent modern strategies present in machine learning research, including:

\begin{itemize}
\item Utilize transfer learning concepts by allowing a front-end model to generate a probabilistic prediction before making a true (regression) prediction. While not directly utilizing transfer learning, the introduction of a predicted probabilistic input vector further informs a regression stage utilizing observed data.

\item The probabilistic classification input vector enables the regression model to adapt to local conditions, ensuring robustness and accuracy across varying reactor states. **This adaptability supports the digital twin’s requirement to operate reliably under diverse scenarios.

\item  By extracting and processing relevant information early, the Pseudo-autoencoded input structure of models ensures computational efficiency and meaningful feature representation, a necessity for digital twins to perform real-time inference and diagnostics effectively.

\end{itemize}

These strategies combine to create a computationally efficient model that is deployable in a digital twin prototype and provides timely, actionable insights to operators and engineers. The model enhances operational efficiency by enabling informed decision-making and presenting additional, interpretable information to personnel in critical moments.

\textbf{Novelties }Introduced in this paper are:
\begin{itemize}
    \item Multi-Stage Deep Learning Framework: A novel framework combining feed-forward neural networks for classification and regression, specifically tailored for predicting reactor power states. This modular, layered design reflects the architecture of digital twins, facilitating synchronization with reactor operations and enabling predictive insights.
    \item Hybrid Data Approach: Integrating real-world reactor measurements with simulated, noise-enhanced data for training ensures robustness and reliability. This hybrid strategy supports the production of a high-fidelity digital twin that can generalize across real-world reactor conditions and respond dynamically to changes.
    \item Dataset Creation and Pre-Processing Techniques: The development of a unique dataset through physical sampling methods and the subsequent pre-processing techniques, including normalization and input selection based on the physical significance of variables, represent significant contributions to the field.
\end{itemize}

\section{Procedure and Multi-Stage DNN Development}
Developing a deep learning model involves developing a pipeline for the data, training, and model development. The following section outlines the development of the pipeline in the standard order, specifically data collection, data processing, and model development. The training and validation results are excluded from this section and instead included in the Results and Discussion.

\subsection{Data Collection}
MSTR is the research nuclear reactor on the campus of Missouri S\&T. This reactor, with a maximum rated power of 200kW, is used frequently for training reactor operators and other personnel compared to other research reactors, such as those focusing on irradiation experiments. This operational diversity provides a dynamic dataset well-suited for developing a digital twin capable of capturing frequent power changes and transient dynamics. This data space can then be utilized to extract additional information regarding power changes in small reactor systems in general. The bulk of power change data available from MSTR includes power changes from operator training and reactor operations courses. The state-based nature of this data aligns with digital twin requirements for accurate system representation and real-time synchronization.

\begin{figure}[htbp]
    \centering
    \includegraphics[scale=0.5]{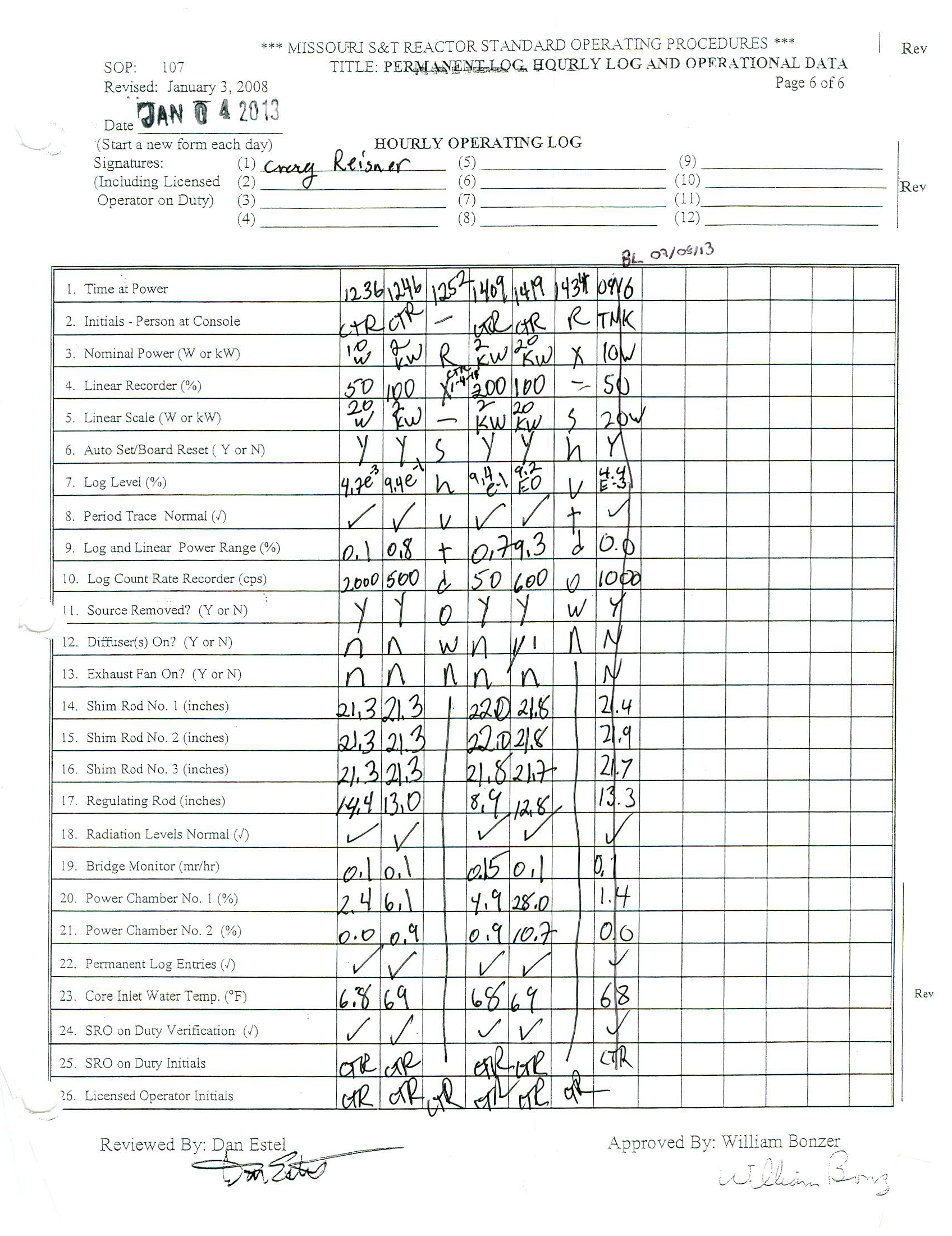}
    \caption{Example MSTR Transient log}
    \label{fig:MSTR_Log}
\end{figure}

Data was collected from MSTR via daily transient logs as shown in Figure \ref{fig:MSTR_Log}. The transient logs provide a foundation for state-based digital twin modeling, where stable reactor power states can define each transient. A single power transient could be defined by generating start and end points based on stable reactor power. These two points in time could then be used to define a state-based approach to the reactor transient, where some information from the initial and final state could be used to make predictions. This state-based approach mirrors the system’s physical behavior and ensures a digital twin has the capability to remain synchronized with real-world operations.

The same information was collected from each stable power, with two stable powers forming a single observation for the state-based prediction model. These stable power observations serve as discrete snapshots of the system's operational state, forming the building blocks for the digital twin. The information from any given stable power could be coupled with the previous or subsequent information from a stable power to create a single data point. The information collected from each of these points included the stable power, control rod heights, date, and time. The digital twin can dynamically predict and simulate the system's future states by retaining this information.

Data could be collected in this format and exported to an Excel file. Each observation included the date, start time, end time, initial power, final power, and initial and final control rod heights. Date and time data were retained for indexing, while the power and control rod information could be used to define the physical state of the reactor system.  This comprehensive data format supports a digital twin's requirement for high-fidelity synchronization with real-world operations. Transient data ranging from January 7, 2013, to October 10, 2014, was collected and processed by hand in this manner.

Some observations were excluded from the collected data. For example, reactor shutdowns were not included due to the excess negative reactivity in the core, which could result in model memorization of shutdowns or negatively impact performance by preferential training towards shutdowns. Transients longer than one hour (from start to finish) were excluded to focus the model on short-term state-based modeling. This focus on short-term transients aligns with the operational needs of a digital twin, ensuring accurate and timely predictions for operational decision-making. After data auditing and exclusion, 542 total observations were collected.

Unfortunately, 542 observations would not be enough to train a model adequately. A physical sampling method was implemented to address the challenge associated with the inadequate training data. The power defect equation, a generalization of the prompt jump equation, was used to generate these additional samples \cite{lamarsh2001introduction}. This equation is generalized as equation \ref{eq:1}. This augmentation process expands the model's training dataset, enabling it to generalize across diverse operational scenarios while preserving fidelity to physical principles and enabling expanded deployment in digital twin-type systems. The power defect was applied to the final state in observation to jump from one stable power to another stable power effectively. The power defect method was performed under the condition that the total reactivity change in this state could not exceed 0.5\$ and no control rod exceeded the 24 inches of maximum withdrawal. This sampling method could be applied $n$ times and form the bulk of the training data while retaining the actual observations for the testing set. The relationship between the reactor power and reactivity can be expressed by Eq. \ref{eq:1}:

\begin{equation}
\label{eq:1}
    \frac{P^{+}}{P^{-}} = \frac{1-\rho^{+}}{1+\rho^{-}}
\end{equation}
where $\rho^{-}$ is lower reactivity, $\rho^{+}$ is higher reactivity, $P^{+}$ is higher power, and 
$P^{-}$ is lower power.

A method was defined using a Python code to randomly perturb control rod heights from both the initial and final states of observation. These perturbations add diversity to the dataset, enhancing the model’s robustness and adaptability to dynamic system changes.  The functions used to execute this practice are included as \ref{sec:power_ratio_sample_func} and \ref{sec:over_sample_func}. Both the initial and final states were perturbed with this method in an effort to reduce the memorization capabilities of the model. The reactivity was determined by using the integral rod worth information for the relevant core configuration based on the date the transient occurred. Over the two years of training samples collected, four different reactor configurations were utilized. The date of the observation could be cross-referenced to the operational dates of the core configurations in order to determine which set of control rod-worth values should be used for the selected sample. For the conversion of the given rod worth values, a delayed neutron fraction ($\beta$) of 0.006 was assumed.

An additional dataset generation method using a joint statistical distribution was performed to increase dataset breadth and smoothness. This method enhances the model’s ability to simulate diverse operational scenarios by generating data with statistical variability and smooth transitions between states, necessary for sufficiently robust operation in digital twins.  A random normal value was generated, which then served as the upper and lower limit to a random normal distribution. This random normal distribution was used to perturb control rod heights and then calculate reactivity insertion. This methodology increases diversity in individual control rod heights and reduces direct clustering around semi-random points in the data. Additional dataset generation methods may be provided upon request to authors.

\begin{figure}[htbp]
    \centering
    \includegraphics[scale=0.6]{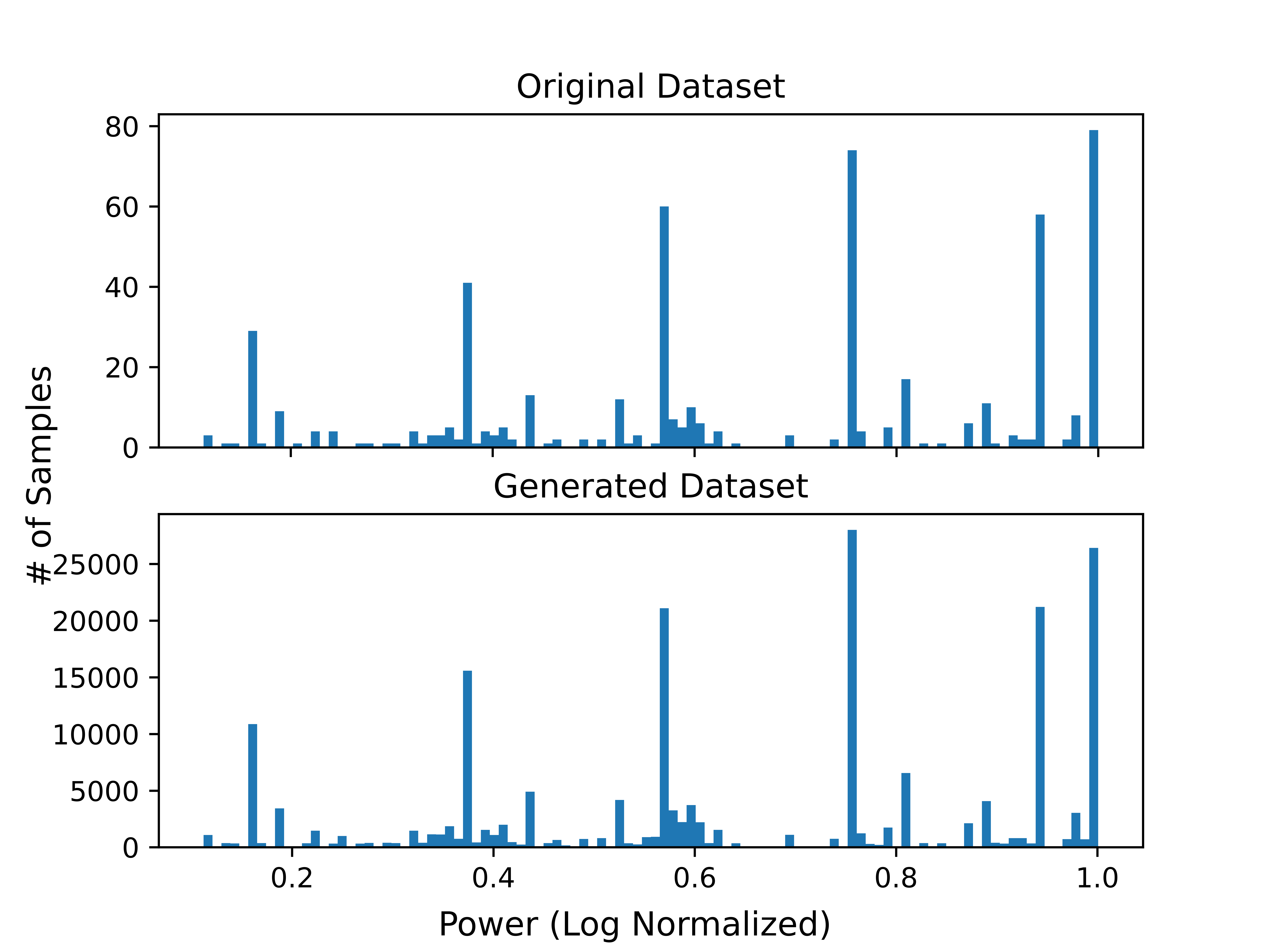}
    \caption{Dataset histograms (top: original, bottom: generated)}
    \label{fig:3}
\end{figure}

Table \ref{tb:rod_worth} represents the relationship between the core configuration and the control rod worths. Using the control rod worths, additional data points could be generated using some random noise. These enriched datasets ensure that the model reflects real-world variability while maintaining high fidelity to reactor physics. The original dataset and generated dataset can be seen in Figure \ref{fig:3}.
% \ref{fig:2} and 

\begin{table}[htbp]
\centering
\caption{Integral Rod Worth for Reactor Core Configurations}
\label{tb:rod_worth}
\begin{tabular}{@{}cllll@{}}
\toprule
\multicolumn{1}{l}{\multirow{2}{*}{Core Configuration}} & \multicolumn{4}{c}{Worth (\$)}  \\ \cmidrule(l){2-5} 
\multicolumn{1}{l}{}                                    & Rod 1 & Rod 2 & Rod 3 & Reg Rod \\ \midrule
120                                                     & 6.387 & 5.380 & 2.963 & 0.488   \\
121                                                     & 6.387 & 5.380 & 2.963 & 0.488   \\
122                                                     & 6.597 & 5.398 & 2.963 & 0.387   \\
123                                                     & 6.583 & 5.267 & 3.017 & 0.433   \\ \bottomrule
\end{tabular}
\end{table}

%%%% Put Figure 3. Dataset histograms (left: original, right: generated)

After samples were generated, processing methods could be applied in order to prepare the data for use in deep learning models.

%%%%
\subsection{Pre-Processing}
Data pre-processing is required for neural network training methods in order to ensure a model is effectively trained, as well as allow the model to retain physical significance and interpretability. For the dataset utilized by this model, pre-processing focused on both data normalization and input/ output determination. Normalization improves the way deep neural network models process input information. Determining the best inputs and outputs for a model governs how information is extracted from it. Both normalization and data selection strategies aim to improve the accuracy and efficacy of the model.

%%%
\subsubsection{Dataset Normalization}
The pre-processing procedure primarily centered around the normalization of the data set and preparation of the target outputs to be used during the training process. A method to normalize a single sample could be developed, extending to the entire dataset, whether actual observation or generated physical sample. Control rod heights were normalized from 0 to 1, with 1 being a full withdrawal of 24 inches. The equation used to normalize control rod height can be seen as equation \ref{eq:3}.

\begin{equation}
\label{eq:3}
    x_{norm} = x_{rod}/24
\end{equation}

Reactor power would need to be normalized and scaled to produce accurate results. To perform this, the natural logarithm of the power was calculated and then normalized to the natural logarithm of full power (200,000 W). The equation used to normalize reactor power can be seen as equation \ref{eq:2}. 

\begin{equation}
\label{eq:2}
    P_{norm} = ln(P)/ln(200,000)
\end{equation}

The single observation normalization method is included as \ref{sec:normalize_sample_func}. This method could then be applied to the entire dataset to form the input data of the deep learning model. The equations used in the normalization process are shown below.

The target outputs could then be generated for the model. With a classification scheme, a series of bins could be used to determine the output. The ceiling of each bin was pre-defined in order to separate the data best. Since outputs tended to cluster around exponentials, the classification method would have to be designed such that similar power changes were still grouped together. The classification categories can be seen in Table \ref{tb:ori_dataset}.

\begin{table}[htbp]
\centering
\caption{Original Dataset 5-class Output Distribution}
\label{tb:ori_dataset}
\begin{tabular}{@{}llllll@{}}
\toprule
Bin \#        & 0  & 1   & 2    & 3     & 4      \\ \midrule
Ceiling (W)   & 90 & 900 & 9000 & 90000 & 200000 \\
\# of samples & 73 & 94  & 100  & 126   & 149    \\ \bottomrule
\end{tabular}
\end{table}

If all samples were properly classified, undersampling methods could be used. Undersampling is important for models which might have much more of one output than another. The model could preferentially select the more populated class resulting in underfitting. An undersampling method was applied to the generated dataset such that all output classes would have the same number of samples. The undersampling method is included as \ref{sec:undersample_from_forted_func}.

%%%
\subsubsection{Input Selection}
After data was generated and normalized, the ideal inputs and outputs could be determined. This was primarily performed by analyzing the physical significance of the variables and determining the best separation methods. Using a state-based classifier, the input variables could be mapped to see what the target class would be before they are used to train the model. Plotting the different variables against each other could show which variables have the most significant separation for the classification method.

Different input variables were posed and compared based on their physical significance. Due to the state-based nature of the network, information from both the initial state and final state would need to be used. Control rod positions were available in both of these states. Control rods are the primary governor of reactivity changes in the system, resulting in those variables being a good choice. With the integral rod worth known, the approximate reactivity contribution from rod withdrawal could be estimated as well. Using control rod heights as an input variable, a model could learn the physical properties of the reactor internally. This could result in a model that predicts a final power state by processing control rod height information about the physical state of the reactor. Alternatively, by using the reactivity insertion provided by the control rod heights, the model could predict the state by attempting to model an equation for reactor power directly.
Another possible input option was initial power. The normalized initial power would be input and could be used by the model to generate an image of the initial state of the system. This would also allow the model to generate ‘jumping off points’ from the common starting states of the reactor. This is due to reactor operators preferentially reaching whole numbers and multiples of 10 when conducting a power change. It could help by clustering the input information with other similar samples.

Along with the clustering from initial power, the direction of change could be input as well. Although the model would hopefully use the differences between initial and final states to predict what the power would be internal, providing information about whether a power increase or decrease is performed could help with clustering. This could be paired with the initial power as well, which would further reinforce the helpful clusters. For instance, by grouping power increases starting at 2 kW separately from power decreases starting at 2 kW, accuracy could be increased. These inputs could be created and separated quickly from the generated dataset and then used to train the deep learning models.

\begin{figure}[htbp]
    \centering
    \includegraphics[scale=0.6]{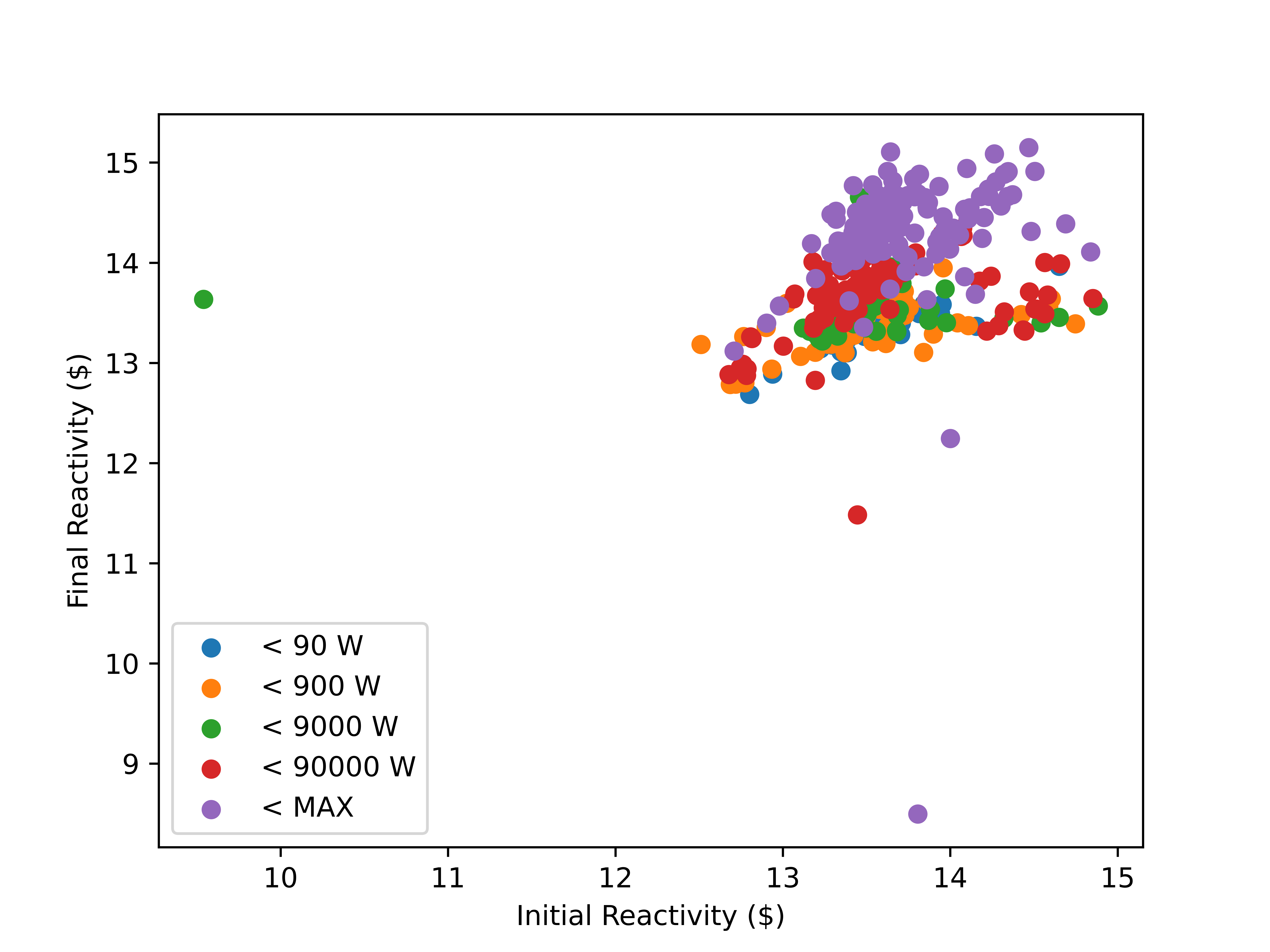}
    \caption{Classification: Reactor power versus inserted reactivity}
    \label{fig:4}
\end{figure}

The datasets split into classes plotted against the relevant input variables can be seen in Figure \ref{fig:4}.
%and Figure \ref{fig:5}.

%%% Put Figure 4. Classification versus rod heights

%%% Put Figure 5. Classification versus reactivity from rods
% \begin{figure}[htbp]
%     \centering
%     \includegraphics[scale=0.6]{image5.png}
%     \caption{Classification: Reactor power versus reactivity from rods}
%     \label{fig:5}
% \end{figure}

\subsection{Training Metrics}
Each model was trained using a loss metric and early stopping evaluated using a separate accuracy metric.

Classification models utilized categorical crossentropy loss and categorical accuracy stopping metric. Categorical accuracy can be calculated using equation \ref{eq:cat_acc}, where $c$ is the number of correct predictions and $N$ is the total number of predictions, calculated over all classes.

\begin{equation}
    ACC = \frac{c}{N}
    \label{eq:cat_acc}
\end{equation}

Regression models utilzed mean absolute error loss and Mean Absolute Percentage Error (MAPE) stopping metric. MAPE can be calculated using equation \ref{eq:mape}, where $p$ is a given predicted value, $t$ is the paired target value, and $N$ is the number of samples.

\begin{equation}
    MAPE = \frac{\sum{\frac{(p - t)}{t}}}{N} * 100
    \label{eq:mape}
\end{equation}

\subsubsection{Neural Network Design}
Multiple deep learning models were generated and trained to determine the best design for predicting energy output. The basic type of network used was a feedforward model, although different ways to input the data, as well as internal structures, were tested. Regularization methods were deployed in every model to reduce overfitting, with L1 and L2 regularization being used most commonly. Models were optimized using either the Adam or Stochastic Gradient Descend (SGD) algorithms. The first models generated were classification networks using categorical cross-entropy loss due to the classification nature of the model. All models were designed using the Keras package of the TensorFlow Python module and trained against the generated data set while the actual collected data was retained as the testing set.

15\% of training samples were retained for a check set for each model. The check dataset was retained to ensure effective training of the model. Specifically, the check set would be run through the model at the end of each epoch. This is so that the neural network will not memorize the original dataset, avoiding overtraining. This is a standard ML checking method from neural network design and is assumed to be an effective training method in this scenario.

Early stopping criterion were developed for the models as well. For classification training, if categorical accuracy was not improved by 0.01\% over 20 epochs, training would halt and the best performing weights would be restored to the model. Similarly, for regression models, an early stopping criterion of 0.1\% Mean Absolute Percentage Error (MAPE) reduction over 20 epochs was imposed. These criterion were performed using the validation dataset to ensure the model would not memorize the dataset. Additional reduction of improvement criteria could be extended to increase the convergence threshold.

Different types of models were designed to see which method would best facilitate the data. This could be typically grouped into two categories: separated input and all-in-one (AIO) input. The AIO input models were formatted like a standard feedforward deep neural network. The input data could be concatenated into a single vector which could be processed by the model for an output. The separated input models attempted to retain some of the physics of the system by having a defined input and output state, which could be processed and then concatenated deeper into the model to be processed for output. A few of the network designs are included as Appendices \ref{app:mi_class_graph} and \ref{app:aio_class_graph}. In these figures, the paired numbers indicate the size of that layer. The question mark is an artifact from the plotting functions in TensorFlow, indicating that the input dimension could be variable. The second number in the pair indicates the output size of the layer, predefined by the architecture. This is true for all subsequent model architecture plots.

% %%%% Put Fig. 6. Multi-input model using power change direction
% \begin{figure}[htbp]
%     \centering
%     \includegraphics[width=15cm]{image6.png}
%     \caption{Multi-input model using power change direction}
%     \label{fig:6}
% \end{figure}

% %%%% Put Fig 7. AIO input model
% \begin{figure}[htbp]
%     \centering
%     \includegraphics[width=10cm]{image7.png}
%     \caption{AIO input model}
%     \label{fig:7}
% \end{figure}

The data input to some models was different than others. Some models used control rod heights as inputs, while some used total reactivity generated by control rods. Some models reduced the total number of input variables by summing the control rod information (when relevant) into a single variable. If the control rod heights were used this way, only control rods 1, 2, and 3 were summed. If the reactivity provided by each control rod was used, all of the rods were summed into a single variable. Table \ref{tb:typs_class_models_trained} shows the types of models designed and trained.

\begin{table}[htbp]
\centering
\caption{Types of Classifier Models Trained}
\label{tb:typs_class_models_trained}
\begin{tabular}{@{}lccccc@{}}
\toprule
Model & \multicolumn{1}{l}{\begin{tabular}[c]{@{}l@{}}Separated\\ Input\end{tabular}} & \multicolumn{1}{l}{AIO}           & \multicolumn{1}{l}{\begin{tabular}[c]{@{}l@{}}Uses\\ Rod Heights\end{tabular}} & \multicolumn{1}{l}{\begin{tabular}[c]{@{}l@{}}Uses\\ Reactivity\end{tabular}} & \multicolumn{1}{l}{Direction}     \\ \midrule
(a1)  & \cellcolor[HTML]{C0C0C0}\checkmark  & \cellcolor[HTML]{C0C0C0}  &   \cellcolor[HTML]{C0C0C0}    & \cellcolor[HTML]{C0C0C0}\checkmark    & \cellcolor[HTML]{C0C0C0}\checkmark \\
(b1)  & \checkmark  &   & \checkmark    &   & \checkmark\\
(c1)  & \cellcolor[HTML]{C0C0C0}\checkmark  & \cellcolor[HTML]{C0C0C0}  & \cellcolor[HTML]{C0C0C0}    & \cellcolor[HTML]{C0C0C0}\checkmark    & \cellcolor[HTML]{C0C0C0}\\
(d1)  & \checkmark  &   & \checkmark    &   &   \\
(e1)  & \cellcolor[HTML]{C0C0C0}    & \cellcolor[HTML]{C0C0C0}\checkmark & \cellcolor[HTML]{C0C0C0}    & \cellcolor[HTML]{C0C0C0}\checkmark    & \cellcolor[HTML]{C0C0C0}\checkmark \\
(f1)  & & \checkmark    & \checkmark    &   & \checkmark\\ \bottomrule
\end{tabular}
\end{table}

Regression models were created to make a final prediction. The probabilistic output of the classifier networks could help form a better basis to distribute the output values in a regression model, and as such, they were used as input. These models utilized a sigmoid output function to squash the output so it could not produce a value greater than 1, which is the normalized maximum power. The architectures were designed similarly to the classifier models, with some having AIO input while others had split input and internal hidden layers. These models were also designed as feedforward neural networks using backpropagation. In both cases, the 5-label output vector from a classifier was used as an additional input. The two general styles are shown in Appendices \ref{app:reg_graph} and \ref{app:aio_reg_graph}, along with the types of models produced in Table \ref{tb:typs_regression}. All regression models utilized the power change direction. All regression models utilized a mean absolute error loss function.

% %%%% Put Fig 8. Split input regression model
% \begin{figure}[htbp]
%     \centering
%     \includegraphics[width=15cm]{image8.png}
%     \caption{Split input regression model}
%     \label{fig:8}
% \end{figure}

% %%%% Put Fig 9. AIO input regression model
% \begin{figure}[htbp]
%     \centering
%     \includegraphics[width=8cm]{image9.png}
%     \caption{AIO input regression model}
%     \label{fig:9}
% \end{figure}

\begin{table}[htbp]
\centering
\caption{Types of Regression Models Trained}
\label{tb:typs_regression}
\begin{tabular}{@{}lcccc@{}}
\toprule
Model & \multicolumn{1}{l}{\begin{tabular}[c]{@{}l@{}}Separated\\ Input\end{tabular}} & \multicolumn{1}{l}{AIO}           & \multicolumn{1}{l}{\begin{tabular}[c]{@{}l@{}}Uses\\ Rod Heights\end{tabular}} & \multicolumn{1}{l}{\begin{tabular}[c]{@{}l@{}}Uses\\ Reactivity\end{tabular}} \\ \midrule
(a2)  & \cellcolor[HTML]{C0C0C0}\checkmark  & \cellcolor[HTML]{C0C0C0}&   \cellcolor[HTML]{C0C0C0}\checkmark  & \cellcolor[HTML]{C0C0C0}\\
(b2)  & \checkmark  &   &   & \checkmark\\
(c2)  & \cellcolor[HTML]{C0C0C0}    & \cellcolor[HTML]{C0C0C0}\checkmark & \cellcolor[HTML]{C0C0C0}\checkmark  & \cellcolor[HTML]{C0C0C0}\\
(d2)  & & \checkmark    &   & \checkmark\\ \bottomrule
\end{tabular}
\end{table}

The model designs proposed in this paper follow the basic framework of feedforward deep neural networks using backpropagation. The complete models are proposed to function in series, in which the results from the classification models feed an additional input similar to keyword inputs from the traditional neural network design. The benefit of feedforward models is their ability to generalize and low computational resource requirements.
The goal of these model designs was to allow for short-term forecasting during power change operations, which could assist reactor operators and allow for operational changes if necessary. Functions were created so each model could be quickly initialized. Once the data set was loaded and the model was initialized, each could be trained.

\subsubsection{Model Hyperparameters}
All models were trained using the ADAM optimizer, with a learning rate of 0.0005. Additionally, L2 regularization was utilized in classification model output layers and L1 regularization was utilized in regression model output layers. L2 regularization was utilized for classification models to avoid reducing connection weights to zero, ensuring the models would still have the capability of making a prediction in any category.

Classification models utilized the softmax output activation to produce a probabilistic output vector. The value of the softmax in a given bin is the predicted effective likelihood that the final reactor power would be in that region. Categorical cross-entropy loss metrics were utilized, with validation using categorical accuracy.

Regression models utilized sigmoid output activation. The sigmoid function is highly non-linear and produces values from 0 to 1. Since reactor power was log-normalized and could not be below 0W power, the sigmoid function would always produce values in the relevant region while extracting non-linear behavior from the previous layer. Mean absolute error loss metrics were utilized, with validation using MAPE.

All hidden layers in the models utilized hyperbolic tangent activation. Hyperbolic tangent is a highly non-linear function with values ranging from -1 to 1. This allows for greater diversification in neuron connection weights and can produce negative correlations in models, such as when making a positive versus negative power change.

Due to the number and nature of the models, the number of neurons is omitted here. Layer units may be seen in the model graphs included in the appendix.

\section{DATA AND RESULTS}
After models were created and inputs were generated, the proposed framework could be used in power prediction. As part of a digital twin framework, these models provide synchronized and accurate representations of reactor states, supporting real-time decision-making and operational optimization. The following sections show the training and testing results, with the results graphed for ease of visualization.

\subsection{Training}
The table below shows the models trained and relevant variables associated with their training. Models were stopped early if the aforementioned early stopping criterion was met. If a model were stopped this way, it would revert to the epoch with the highest validation accuracy. This ensures the trained models retain their performance and robustness for integration into the digital twin, maintaining reliability during operational use. Table \ref{tb:trained_models} includes information from these epochs.

\begin{table}[htbp]
\centering
\caption{Trained models and relevant variables}
\label{tb:trained_models}
\begin{tabular}{@{}llll@{}}
\toprule
Model & Total Epochs               & Best Validation Loss           & Best Validation Accuracy (\%)      \\ \midrule
(a1)  & \cellcolor[HTML]{C0C0C0}71 & \cellcolor[HTML]{C0C0C0}0.384 & \cellcolor[HTML]{C0C0C0}93.1 \\
(b1)  & 64                         & 0.333                         & 95.2                         \\
(c1)  & \cellcolor[HTML]{C0C0C0}92 & \cellcolor[HTML]{C0C0C0}0.335 & \cellcolor[HTML]{C0C0C0}94.3 \\
(d1)  & 60                         & 0.311                         & 95.1                         \\
(e1)  & \cellcolor[HTML]{C0C0C0}66  & \cellcolor[HTML]{C0C0C0}0.462 & \cellcolor[HTML]{C0C0C0}88.4 \\
(f1)  & 67                          & 0.356                         & 93.2                         \\ \bottomrule
\end{tabular}
\end{table}

\begin{figure}[htbp]
    \centering
    \includegraphics[width=13cm]{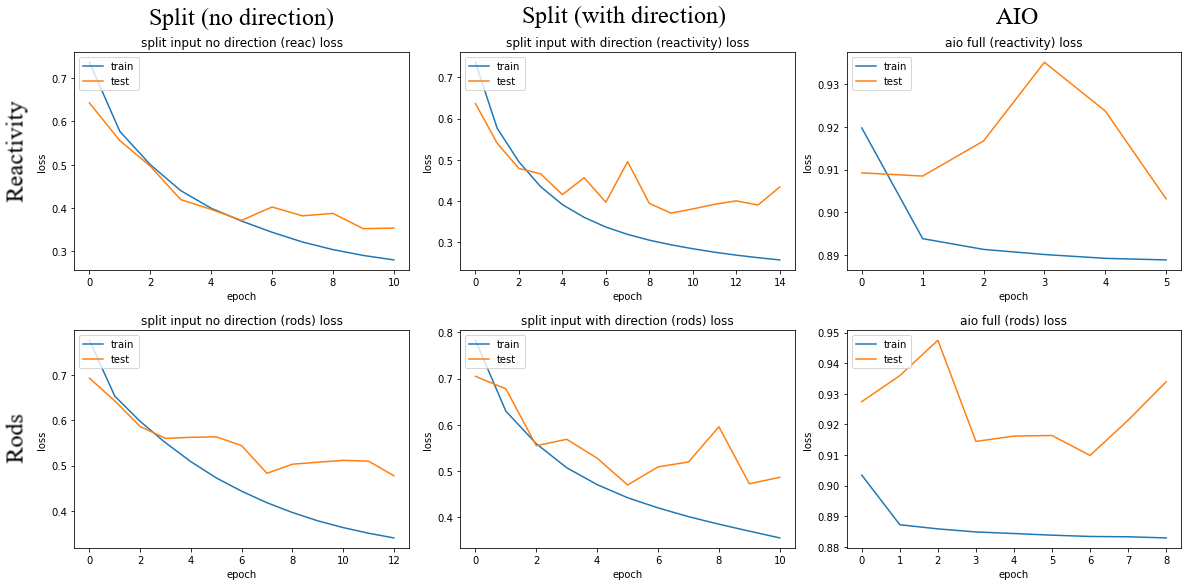}
    \caption{Classifier training losses}
    \label{fig:10}
\end{figure}

\begin{figure}[htbp]
    \centering
    \includegraphics[width=13cm]{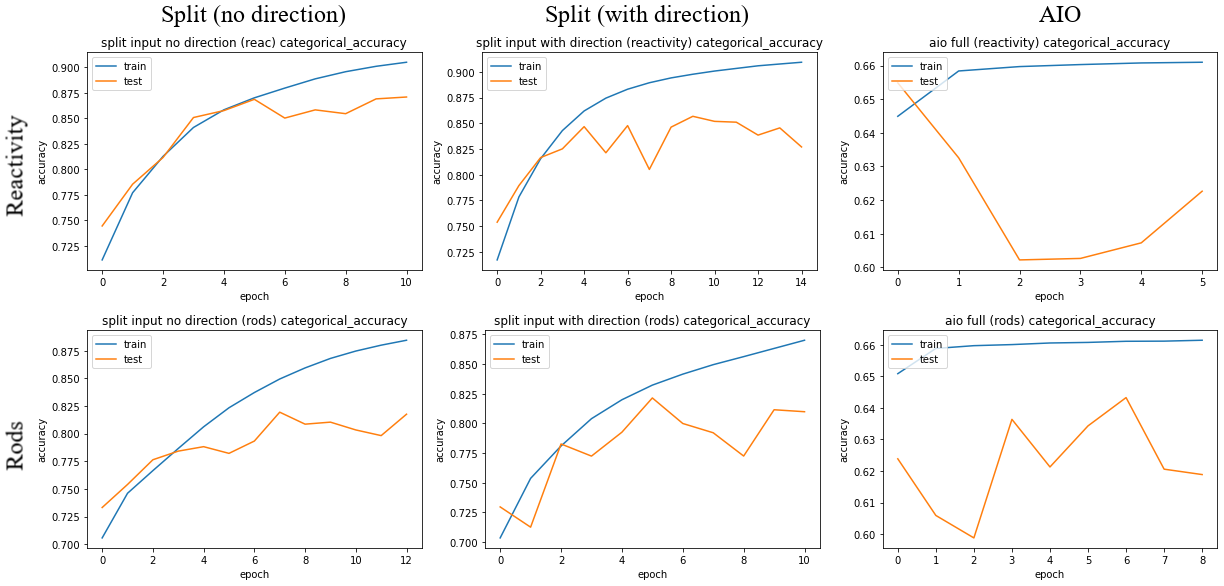}
    \caption{Classifier training categorical accuracy}
    \label{fig:11}
\end{figure}

The loss and accuracy metrics of the models trained are included as Figures \ref{fig:10} and \ref{fig:11}. A history of the training cycle was retained for this purpose. Figure \ref{fig:10} depicts the training losses of each model per epoch of training, and figure \ref{fig:11} depicts the accuracy of each model per epoch trained. These results illustrate the ability of digital twin deployable models to rapidly converge and achieve high accuracy, demonstrating their suitability for real-time deployment. High-performance models were saved in case they would need to be loaded in future instances.

It can be seen that while the split-input type models performed well, the AIO input models trained poorly. This is likely due to a lack of feature separation, causing confusion in the model. This observation demonstrates the importance of preserving physical structure in the input data, a key requirement for maintaining a digital twin's fidelity to the physical reactor system. This would imply that input separation improves training for a physical type of problem.

%%%% Put Figure 10. classifier training losses

%%%% Put Figure 11. classifier training accuracies

Regression models were trained using the output from the classifier most closely resembling its input. Early stopping criteria were applied, with the AIO regression predictors monitoring mean squared error and split input predictors monitoring loss. Models could be tested with the retained data after they were trained. The data from the regression predictor early stopping is included in Table \ref{tb:regression_predictor}.

\begin{table}[htbp]
\centering
\caption{Regression predictor training}
\label{tb:regression_predictor}
\begin{tabular}{@{}llll@{}}
\toprule
Model & Total Epochs               & Best Validation Loss           & Best Validation MAPE (\%)       \\ \midrule
(a2)  & \cellcolor[HTML]{C0C0C0}123 & \cellcolor[HTML]{C0C0C0}0.0317 & \cellcolor[HTML]{C0C0C0}5.277 \\
(b2)  & 126                         & 0.0277                         &  4.082                         \\
(c2)  & \cellcolor[HTML]{C0C0C0}84  & \cellcolor[HTML]{C0C0C0}0.0497 & \cellcolor[HTML]{C0C0C0}9.591 \\
(d2)  & 65                         & 0.0420                         & 7.354                         \\ \bottomrule
\end{tabular}
\end{table}

\subsection{Verification}
The testing results are included below. The goal of the testing is to verify that the models are capable of effective utility in MSTR. It is split into a section for the classifier models and the regression models

\subsubsection{Classifiers}
\begin{figure}[htbp]
    \centering
    \includegraphics[width=15cm]{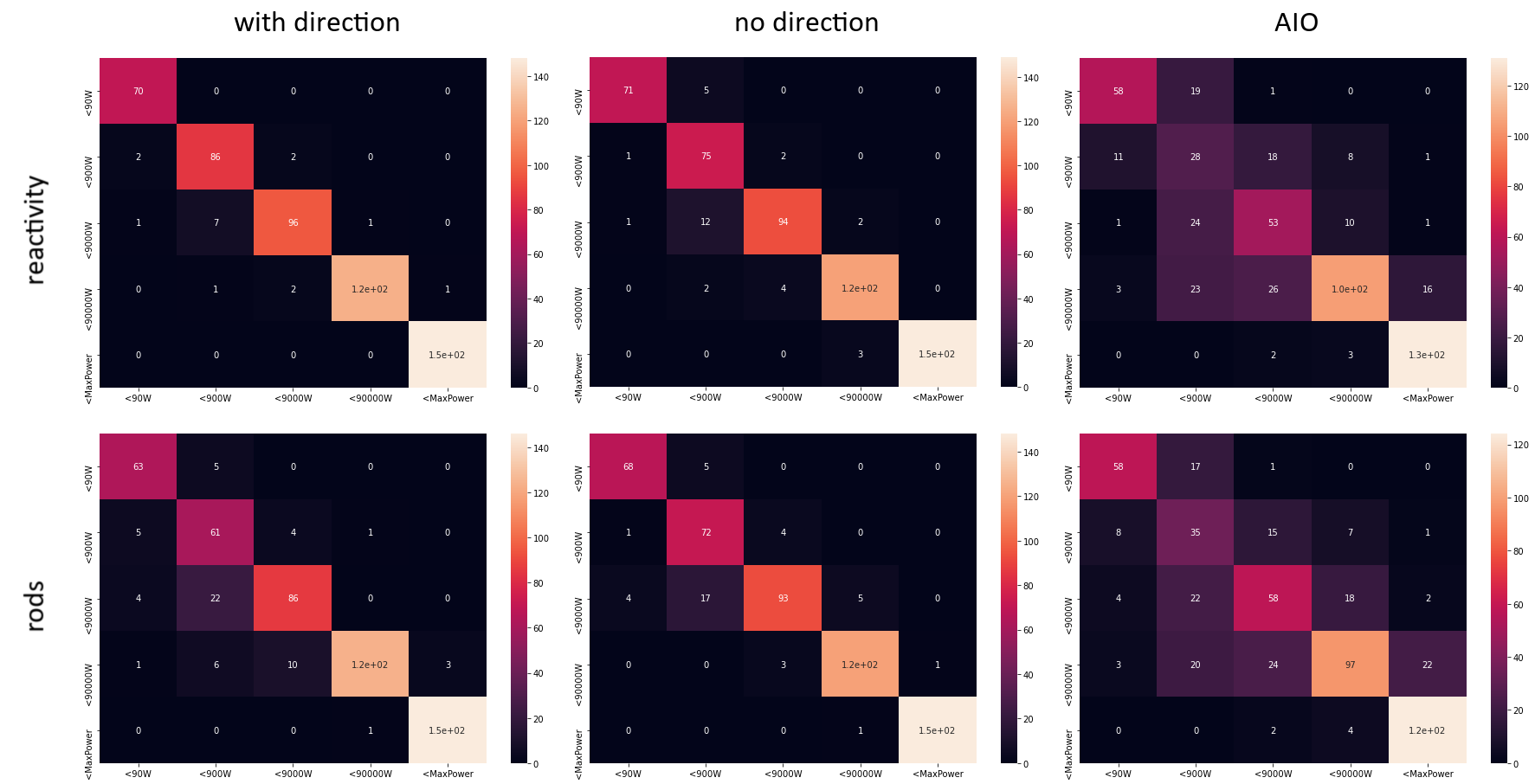}
    \caption{Confusion matrices from recorded data by model}
    \label{fig:12}
\end{figure}

After training, the better performing models could be tested with the actual data from the reactor. Data from the reactor could be formatted into vectors and used by the model to predict outputs. A digital twin may leverage these classifier models to identify operational states and transitions, critical for dynamic synchronization with the physical system. Confusion matrices were used to visualize how well the data performed on the original dataset. The result is shown in Figure \ref{fig:12}.

%%%% Put Figure 12. Confusion matrices from actual dataset

Precision and recall of the models could also be calculated to determine which were the best performers.

\begin{table}[htbp]
\centering
\caption{Classifier model score statistics}
\label{tb:macro_f1}
\begin{tabular}{@{}lll@{}}
\toprule
Model & Categorical Accuracy                      & Macro F1                      \\ \midrule
(a1)  & \cellcolor[HTML]{C0C0C0}0.968 & \cellcolor[HTML]{C0C0C0}0.988 \\
(b1)  & 0.981                         & 0.996                         \\
(c1)  & \cellcolor[HTML]{C0C0C0}0.974 & \cellcolor[HTML]{C0C0C0}0.995 \\
(d1)  & 0.977                         & 0.991                         \\
(e1)  & \cellcolor[HTML]{C0C0C0}0.920 & \cellcolor[HTML]{C0C0C0}0.991 \\
(f1)  & 0.968                        & 0.991                         \\ \bottomrule
\end{tabular}
\end{table}

%%%%%%%%% internal accuracy metrics of models a1, d1, and e1
It can be seen that the reactivity models produced better results than the rod height models on the high end. The internal accuracy metrics of some of the models are listed in Tables \ref{tb:internal_accuracy_a1}, \ref{tb:internal_acuracy_d1}, and \ref{tb:internal_acuracy_e1}.

\begin{table}[htbp]
\centering
\caption{Internal accuracy metrics of model a1}
\label{tb:internal_accuracy_a1}
\begin{tabular}{@{}llll@{}}
\toprule
Class             & Precision                     & Recall                        & F1                            \\ \midrule
\textless{}90W    & \cellcolor[HTML]{C0C0C0}0.985 & \cellcolor[HTML]{C0C0C0}0.970 & \cellcolor[HTML]{C0C0C0}0.978 \\
\textless{}900W   & 0.988                         & 0.927                         & 0.957                         \\
\textless{}9000W  & \cellcolor[HTML]{C0C0C0}0.906 & \cellcolor[HTML]{C0C0C0}0.974 & \cellcolor[HTML]{C0C0C0}0.939 \\
\textless{}90000W & 0.968                         & 0.990                         & 0.979                         \\
\textless{}MAX    & \cellcolor[HTML]{C0C0C0}0.993 & \cellcolor[HTML]{C0C0C0}0.983 & \cellcolor[HTML]{C0C0C0}0.988 \\ \bottomrule
\end{tabular}
\end{table}

\begin{table}[htbp]
\centering
\caption{Internal accuracy metrics of model d1}
\label{tb:internal_acuracy_d1}
\begin{tabular}{@{}llll@{}}
\toprule
Class             & Precision                     & Recall                        & F1                            \\ \midrule
\textless{}90W    & \cellcolor[HTML]{C0C0C0}0.986 & \cellcolor[HTML]{C0C0C0}0.989 & \cellcolor[HTML]{C0C0C0}0.987 \\
\textless{}900W   & 0.958                         & 0.975                         & 0.966                         \\
\textless{}9000W  & \cellcolor[HTML]{C0C0C0}0.969 & \cellcolor[HTML]{C0C0C0}0.946  & \cellcolor[HTML]{C0C0C0}0.957 \\
\textless{}90000W & 0.968                         & 0.989                         & 0.979                         \\
\textless{}MAX    & \cellcolor[HTML]{C0C0C0}1.0 & \cellcolor[HTML]{C0C0C0}0.982 & \cellcolor[HTML]{C0C0C0}0.991 \\ \bottomrule
\end{tabular}
\end{table}

\begin{table}[htbp]
\centering
\caption{Internal accuracy metrics of model e1}
\label{tb:internal_acuracy_e1}
\begin{tabular}{@{}llll@{}}
\toprule
Class             & Precision                     & Recall                        & F1                            \\ \midrule
\textless{}90W    & \cellcolor[HTML]{C0C0C0}0.933 & \cellcolor[HTML]{C0C0C0}0.973 & \cellcolor[HTML]{C0C0C0}0.952 \\
\textless{}900W   & 0.957                         & 0.797                         & 0.870                         \\
\textless{}9000W  & \cellcolor[HTML]{C0C0C0}0.786 & \cellcolor[HTML]{C0C0C0}0.907 & \cellcolor[HTML]{C0C0C0}0.842 \\
\textless{}90000W & 0.923                         & 0.964                         & 0.943                         \\
\textless{}MAX    & \cellcolor[HTML]{C0C0C0}1.0 & \cellcolor[HTML]{C0C0C0}0.984 & \cellcolor[HTML]{C0C0C0}0.991 \\ \bottomrule
\end{tabular}
\end{table}
%%%%%%%%%%%%%%

\subsubsection{Regression}
\begin{figure}[htbp]
    \centering
    \includegraphics[width=11cm]{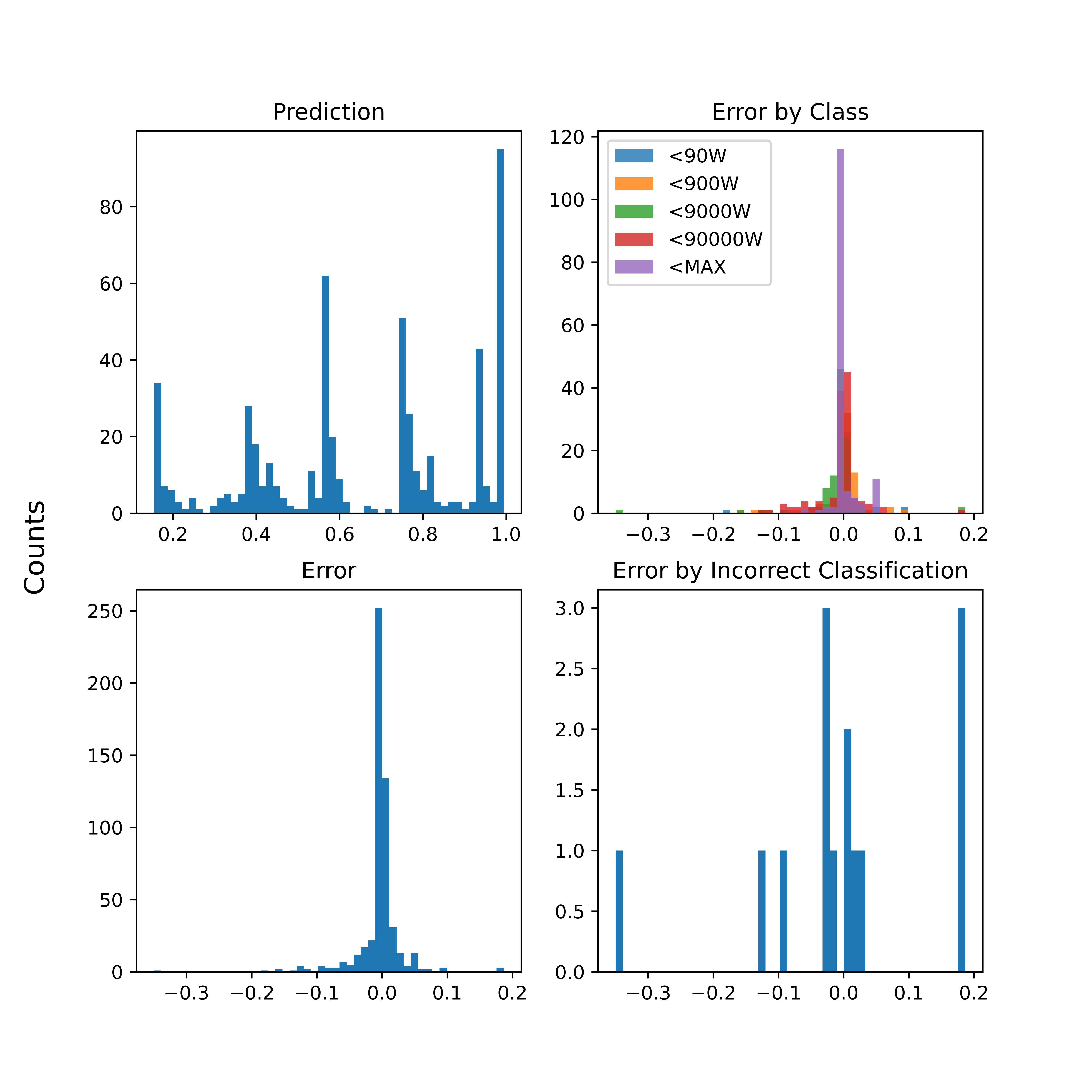}
    \caption{Multi-Input rod data regression predictions and error}
    \label{fig:13}
\end{figure}

\begin{figure}[htbp]
    \centering
    \includegraphics[width=11cm]{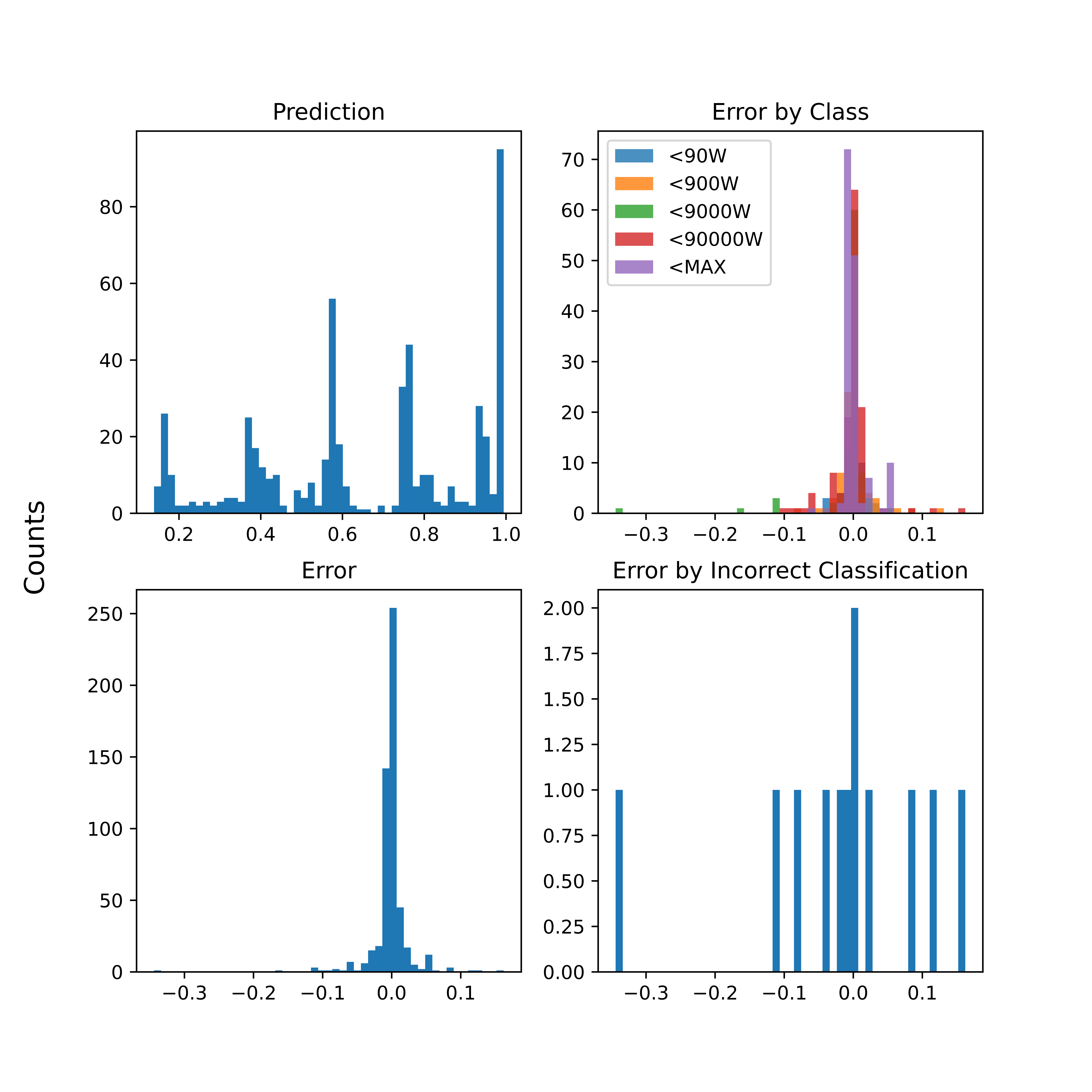}
    \caption{Multi-Input reactivity data regression predictions and error}
    \label{fig:14}
\end{figure}

\begin{figure}[htbp]
    \centering
    \includegraphics[width=11cm]{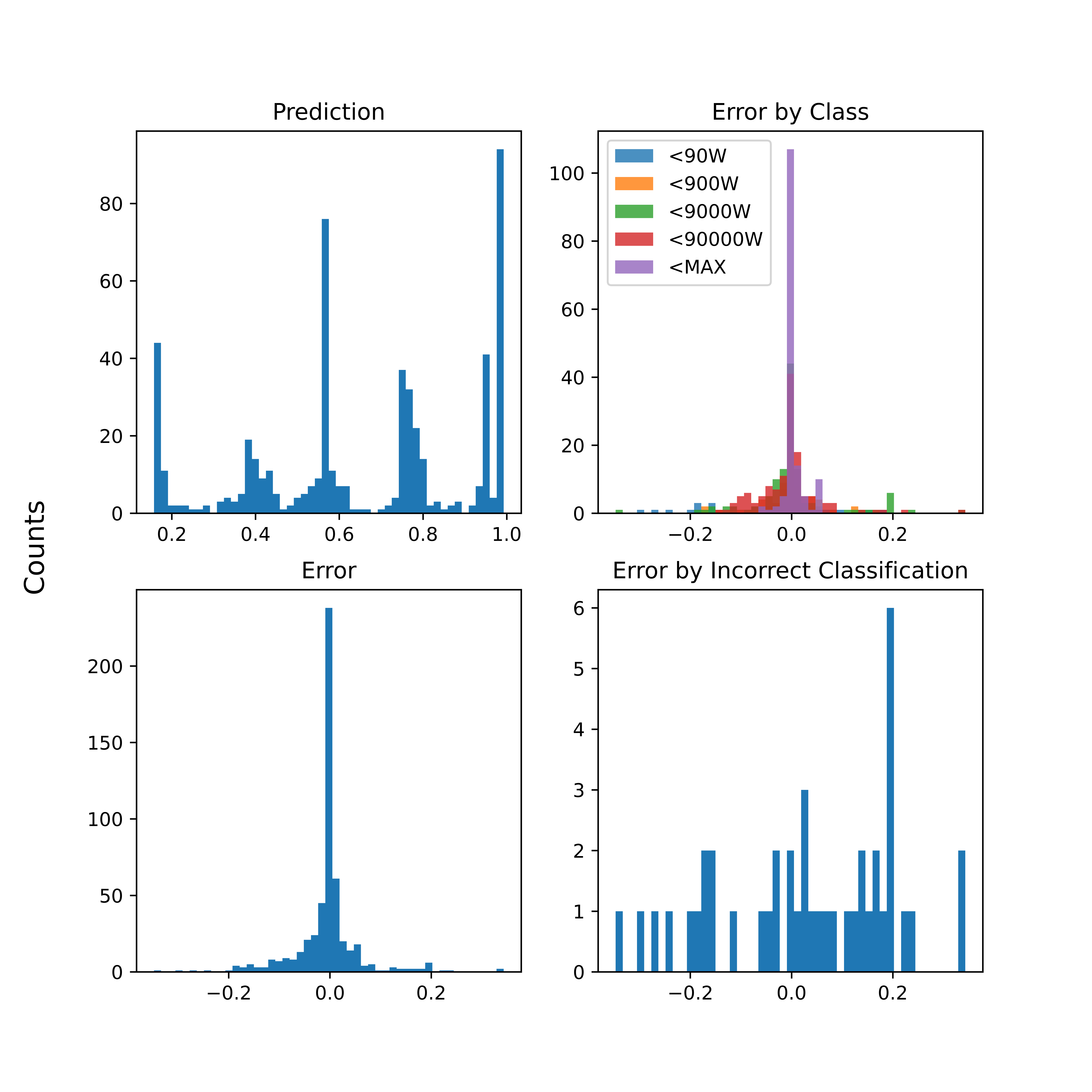}
    \caption{AIO rod data regression predictions and error}
    \label{fig:15}
\end{figure}

The regression models were trained using the output from the classifier most closely resembling its input type. These regression models enable the digital twin to predict reactor power dynamics with high accuracy, providing actionable insights for operational decision-making. The distribution and error of the regression models can be analyzed to show their efficacy. The results are shown in Figures \ref{fig:13}, \ref{fig:14}, and \ref{fig:15}.

%%%% Put Figure 13. reg_full_rods prediction error

%%%% Put Figure 14. reg_full_reacs prediction error

%%%% PUt Figure 15. reg_w_pred prediction error

%%%% Put Figure 17. reg_pred_reacs prediction error
\begin{figure}[htbp]
    \centering
    \includegraphics[width=11cm]{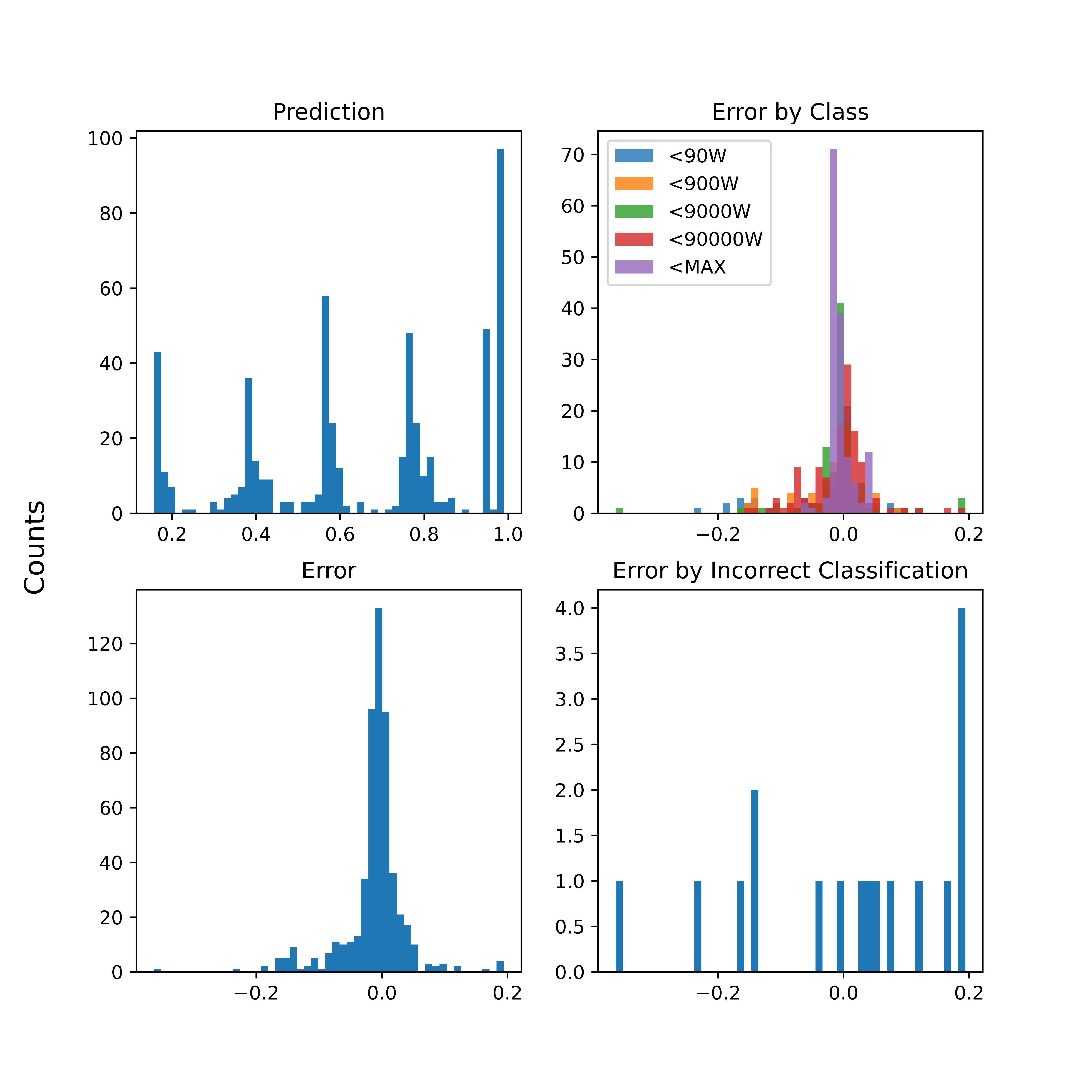}
    \caption{AIO reactivity data regression predictions and error}
    \label{fig:16}
\end{figure}

The results show that the models using control rod height rather than reactivity have a blind spot around the 0.4 output spike present in the original dataset. This could be due to either the worse performance of the first stage classifiers using control rod height or because the model cannot correctly infer the reactor physics from just the control rod heights. The AIO and split input models seem to have only minimal difference between the two, implying that most of the prediction information is inferences from the classifier from the previous stage. Additional regularization of the classification input could help reduce this reliance.

Additionally, the MAPE for each model is examined using the MSTR dataset. Results can be seen in Table \ref{tb:mape_results}.

\begin{table}[htbp]
\centering
\caption{Regression model MAPE}
\label{tb:mape_results}
\begin{tabular}{@{}llll@{}}
\toprule
Model             & MAPE\\ \midrule
\cellcolor[HTML]{C0C0C0}a2    & \cellcolor[HTML]{C0C0C0}3.22 \\
b2   &2.32  \\
\cellcolor[HTML]{C0C0C0}c2  & \cellcolor[HTML]{C0C0C0}7.72  \\
d2 & 5.77 \\ \bottomrule
\end{tabular}
\end{table}

\subsubsection{Combined Model}
\begin{figure}[htbp]
    \centering
    \includegraphics[width=15cm]{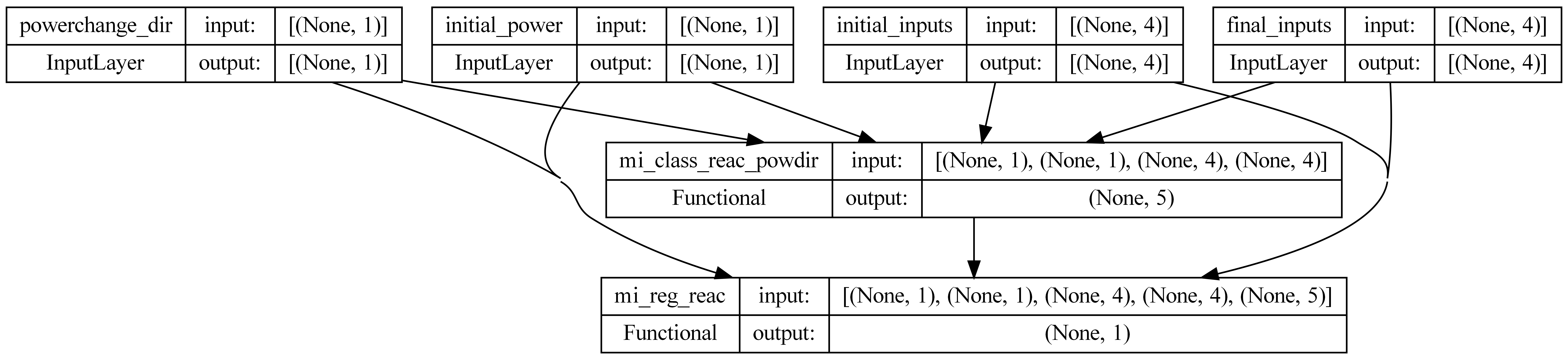}
    \caption{Final two-stage model}
    \label{fig:17}
\end{figure}

To complete the model, the most successful models from each stage can be combined. This integration represents the final step in constructing a digital twin that provides both categorical and continuous predictions of reactor states. A model can be defined in Tensorflow using a trained model from each stage as a functional layer. The result is a model that takes the input utilized in the first stage of the model and internally processes it for use in the second stage of the model. The output of the first stage is also utilized for the input in the second stage, as outlined in the training section. The full model with two functional layers and an input layer can be seen in Figure \ref{fig:17}. The selected first-stage model was a1, and the selected second-stage model was b2.

%%%% Put Figure 17. Final two-stage model

This final model embodies digital twin supporting modules by providing synchronized and accurate predictions of reactor behavior, enabling real-time monitoring, predictive diagnostics, and operational optimization. The resultant DNN outputs both the categorical and numerical predictions. Since DNNs are replicable, the predictions in each stage are the same as the predictions from models a1 and b2. This final DNN can be functionally applied or retrofitted and exported to function as a component in a larger system.

\section{DISCUSSION}
Practically, a similar type of model could likely be developed in other reactors.  As part of a digital twin framework, these models could serve as adaptable virtual replicas tailored to the unique design of each reactor core. The inputs would have to be developed based on the design of the reactor core, and the inputs could further be expanded as well. The normalization of control rod heights could cause the models' reduced accuracy using control rod heights as input. More likely, however, is that the model needs more feature representation to infer the physics of the reactor core. In a digital twin, the integration of reactivity as an input eliminates this limitation by directly linking the reactor power to reactivity, thereby enhancing the twin's ability to mirror reactor behavior accurately.  This allows future models to account for negative reactivity provided by burnable poisons or testing samples. Furthermore, the use of reactivity enables the digital twin to generalize across different reactor core configurations, ensuring its applicability beyond the conditions on which it was originally trained. That said, classifier accuracy for control rods was not impractically low, and most incorrect predictions were in an adjacent energy resolution. The classification accuracy using control rods is at least a proof of concept for future models using similar inputs.

Over the time the data was collected, MSTR saw significant use. This could have led to changes in the core physics. This time period corresponds to a total burnup of 120MWd/MTU. Based on the fuel designs, this is 13\% of the fuel's lifetime burnup. These evolving reactor conditions highlight the importance of dynamic adaptability within a digital twin framework. Future iterations of these types of models in digital twin systems could incorporate support modules that account for changes in core physics due to burnup, fuel poisoning, or operational conditions, maintaining accuracy over the reactor's lifecycle.

Future work would likely take one of two forms: improved data or continued development. Increasing the amount of available data would be useful, but more specifically, increasing the available input features and generating more distributed data would yield better results. Since the power output clusters around the 2*10x powers, the regression model can rely heavily on the classifier output. More distributed power output could force the regression network to train more using the physical information from the reactor. Additional input information, such as the total reactor uptime, presence of a sample, burnable poison, etc., could also be useful. Additional input features could have an effect on accuracy.

The other subject of future research would be continued development. Enhancing these models for further deployment in digital twin systems could involve exploring advanced architectures, such as Long Short-Term Memory (LSTM) networks or convolutional neural networks (CNNs), for dynamic time series prediction. A state-based predictor like the one developed could be used in conjunction with this type of model, which would provide an end state as a horizontal asymptote for the time series. This integration would allow the digital twin to simulate reactor transitions dynamically while also providing an estimated time to reach the final state, further aligning with real-time operational needs. If an automatic controller were desired, one could be developed in conjunction with a state-based predictor as a sort of look-ahead function. Such a controller, integrated with the digital twin, could predict reactor state transitions in real time and provide operational feedback to optimize control actions. The controller would perform an action, and the predictor could generate the final power of the transient and provide feedback to the controller. This modularization would allow the digital twin to operate as an online, automated monitoring and control system, dynamically adapting to live reactor conditions and supporting operational decision-making in real time. In fact, using a similar model to approximate feedback terms from reactor operation could be used in conjunction with other types of models (machine learning or physics-based) to streamline the prediction process. The predictor controller could then be modularized and actively receive live information from the reactor to form an online automatic controller.

\section{CONCLUSIONS \& FUTURE WORKS}
The analysis of the generated models reveals that the classifier models, particularly those leveraging reactivity inputs, exhibit exceptional performance. These models align closely with the principles of a digital twin by providing real-time, probabilistic insights into reactor state transitions, enabling dynamic synchronization with the physical system. The probabilistic nature of the softmax output function facilitates a nuanced evaluation of the model's predictions, further enhancing its practical utility. The preference for reactivity-based models over those utilizing control rod heights underscores the importance of integrating physically interpretable parameters, a key requirement for ensuring the fidelity of digital twin representations. While the inclusion of power change direction yielded marginal gains in classifier accuracy, its omission in future iterations can potentially simplify the framework without compromising performance. The regression models also demonstrated commendable accuracy, with predictions based on accurate classifier outputs exhibiting a minimum MAPE of 2.3\%. This performance showcases the framework's potential as a reliable predictive component within a digital twin, supporting operational decisions with high confidence. The observed increase in prediction error near the maximum power level, attributed to the logarithmic normalization technique, and the occasional 20\%  errors in regression models, potentially linked to power state clustering, highlight areas for future refinement. These insights emphasize the need for continuous updates and calibration, consistent with the adaptive nature of digital twins. The insights gleaned from this study lay a robust groundwork for applying deep learning in nuclear reactor modeling and control, paving the way for developing intelligent digital twin systems that promise advancements in operational efficiency and safety.

The continuation of this study will contribute to the development of an intelligent digital twin by integrating advanced deep learning frameworks to enable real-time monitoring and optimization. The proposed digital twin framework aims to not only replicate but also enhance reactor operations through predictive analytics, real-time diagnostics, and adaptive optimization. Future iterations of the proposed framework could incorporate:

\begin{itemize}
    \item Emerging neural operator methodologies, such as Deep Neural Operators (DNOs) and Fourier Neural Operators (FNOs), to enhance the model’s ability to generalize across varying reactor configurations while reducing computational overhead. 
    
     \item  A robust uncertainty quantification (UQ) framework (derived from authors' previous studies \cite{kobayashi2024ai,kumar2019influence, kumar2022multi}) is essential for ensuring reliable predictions. This could involve Monte Carlo simulations or Bayesian techniques to account for uncertainties in both real-world reactor data and machine learning models.

    \item Integration of DNOs with reactor power prediction models, real-time signal processing \cite{kabir2010theory,kabir2010non,kabir2010watermarking}, and advanced data augmentation techniques for optimizing nuclear reactor operations.
\end{itemize}

\pagebreak

\section*{Appendix}

\subsection{Multi-Input Classifier Model Graph}
\label{app:mi_class_graph}
\begin{figure}[h!tbp]
    \centering
    \includegraphics[width=16cm,angle=270]{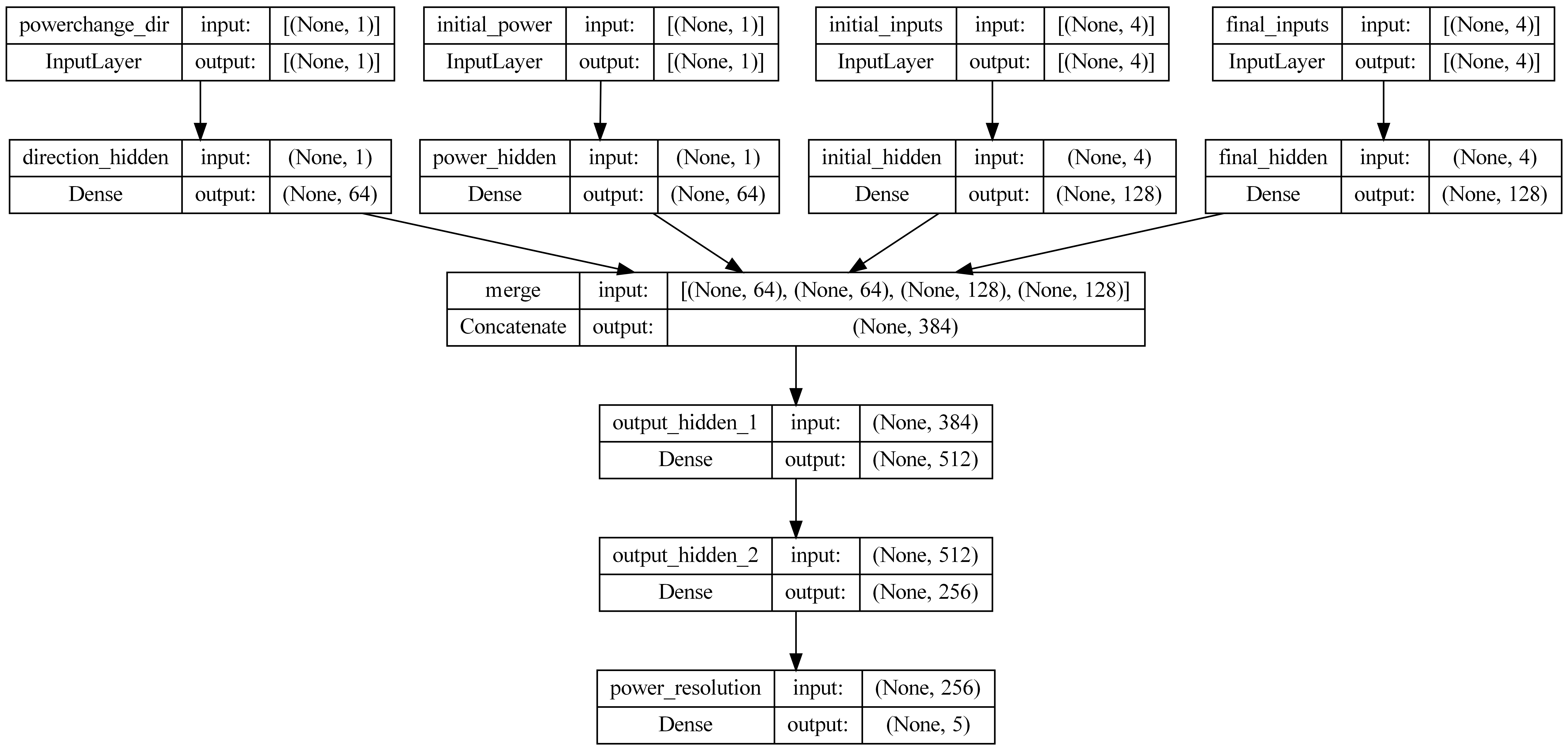}
    \caption{Multi-Input classifier utilizing power change direction model graph}
    \label{fig:6}
\end{figure}

\pagebreak

\subsection{AIO Classifier Model Graph}
\label{app:aio_class_graph}

\begin{figure}[h!tbp]
    \centering
    \includegraphics[width=0.5\linewidth]{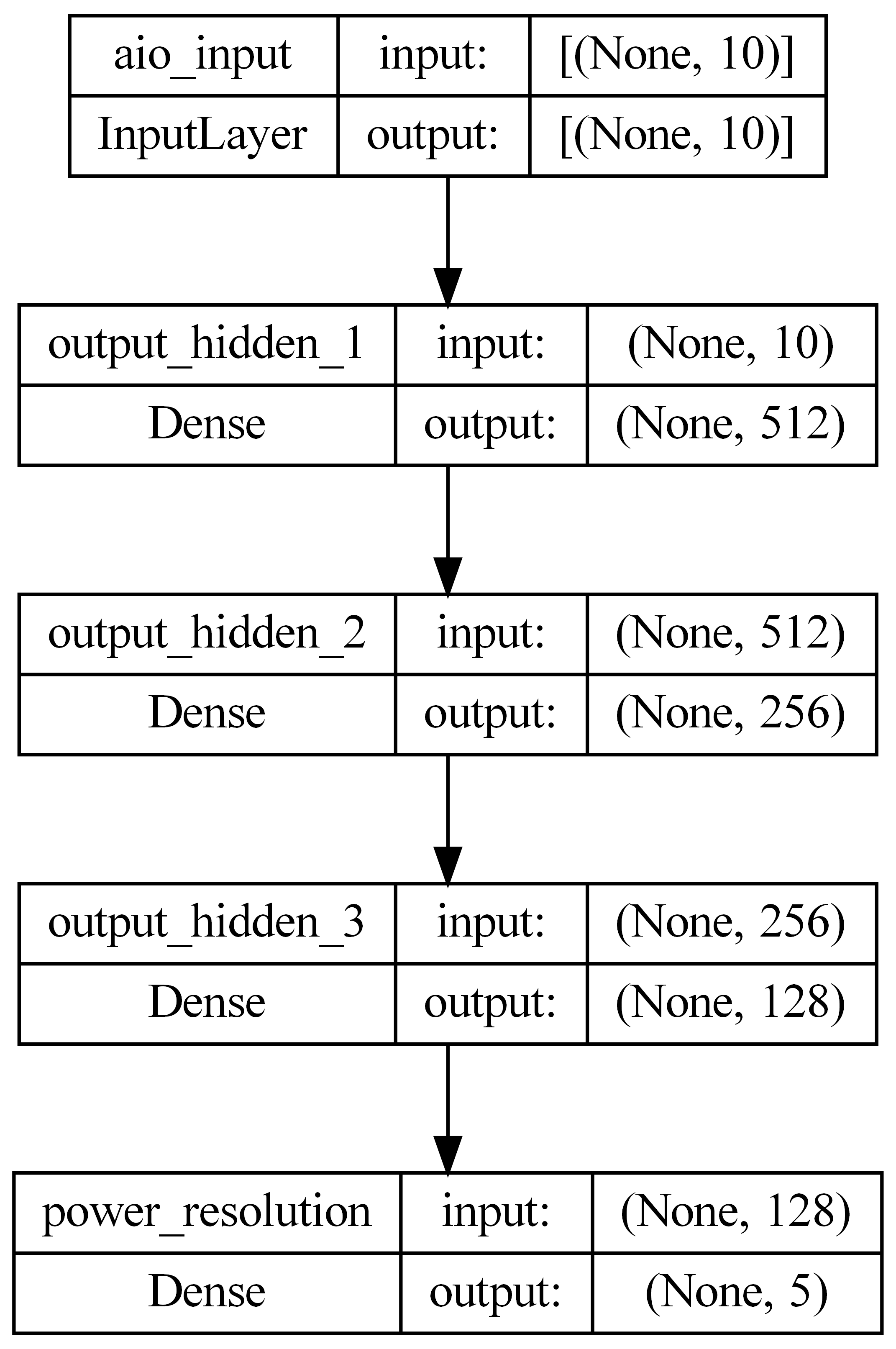}
    \caption{AIO classifier model graph}
    \label{fig:7}
\end{figure}

\pagebreak

\subsection{Multi-Input Regressor Model Graph}
\label{app:reg_graph}

\begin{figure}[h!tbp]
    \centering
    \includegraphics[width=16cm,angle=270]{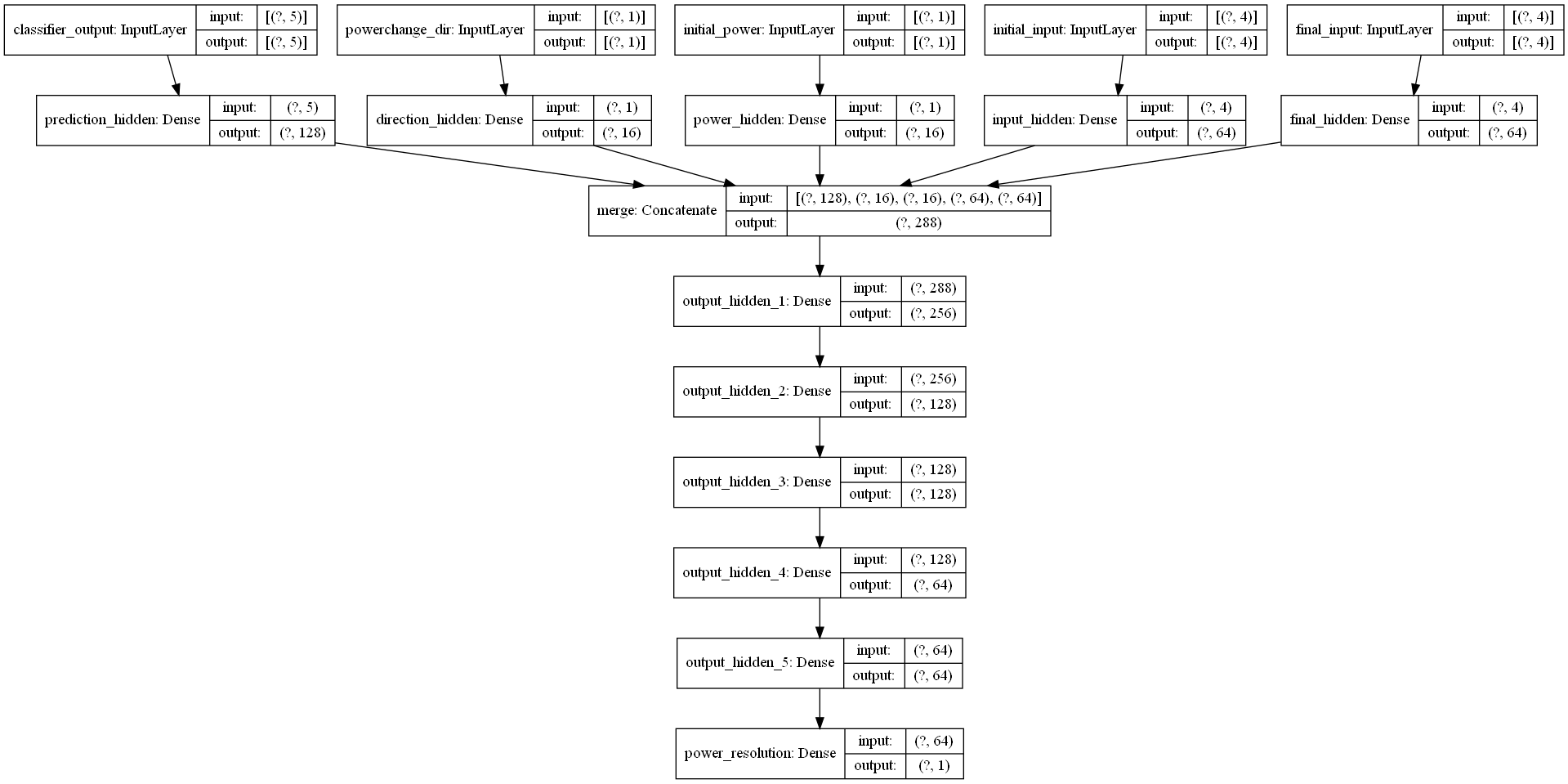}
    \caption{Multi-Input regressor model graph}
    \label{fig:8}
\end{figure}

\pagebreak

\subsection{AIO Regressor Model Graph}
\label{app:aio_reg_graph}

\begin{figure}[h!tbp]
    \centering
    \includegraphics[width=10cm]{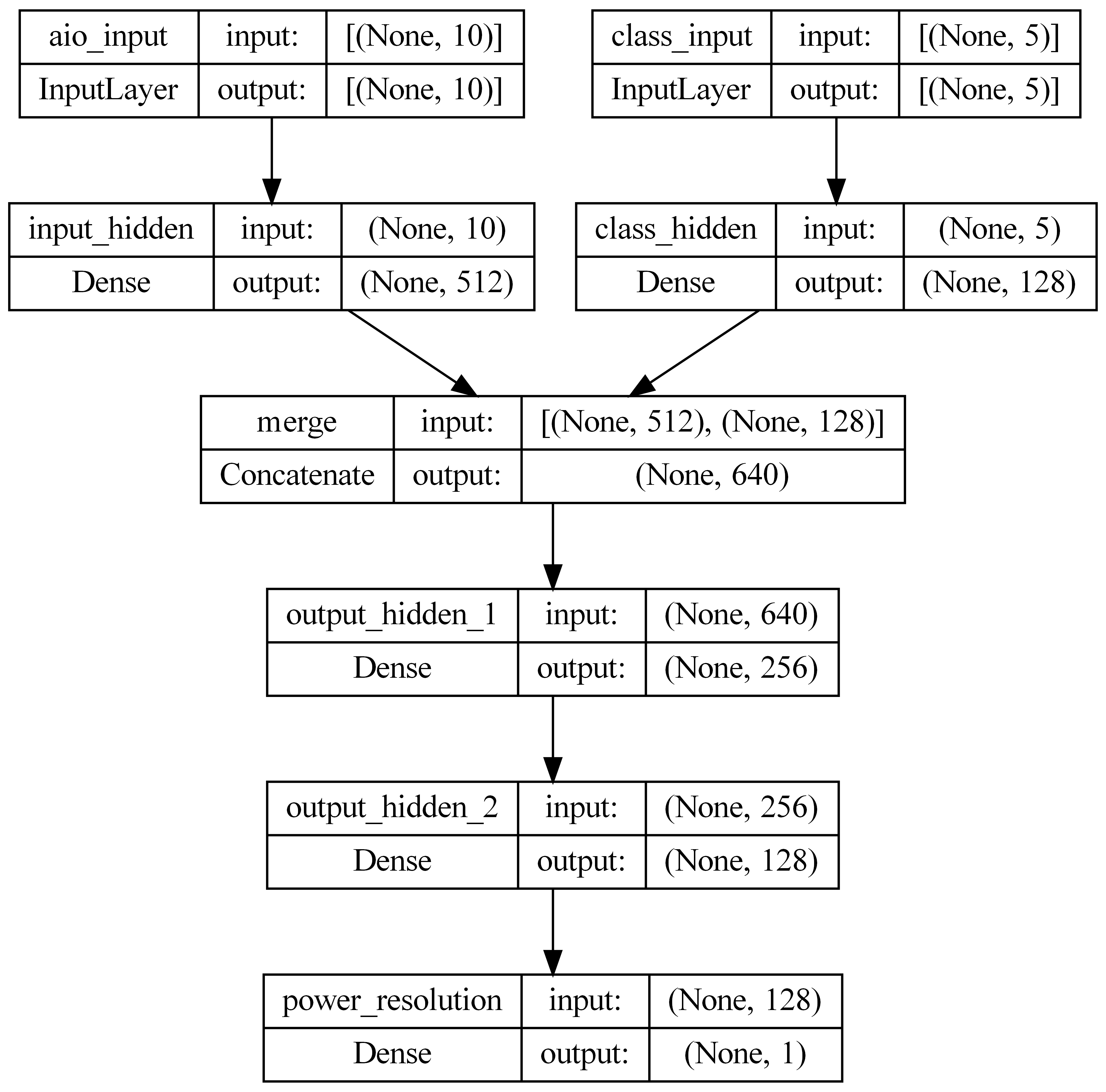}
    \caption{AIO regressor model graph}
    \label{fig:9}
\end{figure}

\pagebreak

\subsection{power\_ratio\_sample function}
\label{sec:power_ratio_sample_func}
\begin{lstlisting}[language=Python]
# creates an additional datapoint from sample
# uses power ratio for small reactivities
# equation from Lamarsh 346-347
# performes microscopic power change on previously existing sample
# assume given 4 rods and reactivities
# allows additional preturbation of regulating rod (pos4)
def power_ratio_sample(rods, power, reacs, change):
    # determines whether the additional power change is positive or negative
    # mainly changes the behavior of the random number generator
    clear=False
    delta = np.random.normal(loc=0.0, scale=0.15, size=4)
    if change==1:
        delta[0] = delta[0]+0.3
        delta[1] = delta[1]+0.3
        delta[2] = delta[2]+0.3
    if change==-1:
        delta[0] = delta[0]-0.3
        delta[1] = delta[1]-0.3
        delta[2] = delta[2]-0.3
    if change==0:
        delta[0] = (delta[0])
        delta[1] = (delta[1])
        delta[2] = (delta[2])
    # preturb regulating rod separately
    delta[3] = delta[3]*10
    new_rods = delta+rods
    # check if all control rod heights are real
    if np.max(new_rods)>24 or np.min(new_rods)<0:
        clear=False
    else:
        clear=True
    # determine initial reactivity
    init_reacs = rods*reacs
    reac_i = np.sum(init_reacs)
    # determine change in reactivity
    delta_reacs = new_rods*reacs
    reac_f = np.sum(delta_reacs)
    # determine micro power change
    new_power = power*((1-reac_f)/(1-reac_i))
    # catch overpowers
    if new_power>200000:
        clear=False3
    # equation becomes invalid for reactivities larger than 0.5$
    if abs(reac_f-reac_i)>0.5:
        clear=False
    # outputs new rod and power state
    return new_rods, new_power, clear

\end{lstlisting}

%%%
\subsection{over\_sample function}
\label{sec:over_sample_func}
\begin{lstlisting}[language=Python]
# creates additional samples en masse
# integral reactivity represented in $
def over_sample(dataset, dates, n, change):
    # create new array for generated data
    new_data = np.empty(shape=(n,11))
    
    # reactivities of each core configuration
    reacs120 = np.array([.03832, .03228, .01778, .00293])/0.006
    reacs121 = np.array([.03832, .03228, .01778, .00293])/0.006
    reacs122 = np.array([.03958, .03239, .01778, .00232])/0.006
    reacs123 = np.array([.0395, .0316, .0181, .0026])/0.006
    # start dates of each core configuration, in ordinal
    core121=date(year=2014, month=1, day=15)
    core121=int(core121.toordinal())
    core122=date(year=2014, month=10, day=9)
    core122=int(core122.toordinal())
    core123=date(year=2014, month=10, day=16)
    core123=int(core123.toordinal())
    
    # generate new samples via power ratio
    for i in range(n):
        # determine which original sample to use
        og = np.random.randint(low=0, high=dataset.shape[0])
        power_i = dataset[og,0]
        power_f = dataset[og,1]
        rods_i = dataset[og,2:6]
        rods_f = dataset[og,6:10]
        date_code = dataset[og,10]
        # get the reactivity for this power change
        day = dates[og]
        if day<core121:
            reacs=reacs120
            config = 0
        elif day<core122:
            reacs=reacs121
            config = 1
        elif day<core123:
            reacs=reacs122
            config = 2
        else:
            reacs=reacs123
            config = 3
        # generate new initial state
        new_rods_i, new_power_i, pass_i = power_ratio_sample(rods_i, power_i, reacs, change)
        # generate new final state
        new_rods_f, new_power_f, pass_f = power_ratio_sample(rods_f, power_f, reacs, change)
        # port new information to output array
        new_data[i,0] = new_power_i # initial power
        new_data[i,1] = new_power_f # final power
        new_data[i,2:6] = new_rods_i # initial rod heights
        new_data[i,6:10] = new_rods_f # final rod heights
        new_data[i,10] = date_code # ordinal date
        # if sample doesn't pass, redo sample
        if not(pass_i) or not(pass_f):
            i=i-1
    return new_data

\end{lstlisting}

%%%
\subsection{normalize\_sample function}
\label{sec:normalize_sample_func}
\begin{lstlisting}[language=Python]
# normalizes one sample for input to models
def normalize_sample(sample, y):
    # normalize inputs
    output = np.empty(shape=(np.shape(sample)))
    output[0] = np.log(sample[0])/np.log(200000) # gets log of initial power
    output[1] = np.log(sample[1])/np.log(200000) # gets log of final power
    output[2:6] = sample[2:6] / 24 # normalizes control rod height to max withdrawal
    output[6:10] = sample[6:10] / 24 # normalizes control rod height to max withdrawal
    output[10] = sample[10]
    # normalize outputs
    y_out = np.empty(shape=(np.shape(y)))
    y_out[0] = y[0]
    if y[1]<0:
        out = np.abs(y[1])
        out = np.log(out)
        out = -out
        y_out[1] = out
    else:
        out = y[1]
        out = np.log(out)
        y_out[1] = out
    return output, y_out
\end{lstlisting}

\subsection{undersample\_from\_sorted function}
\label{sec:undersample_from_forted_func}
\begin{lstlisting}[language=Python]
# takes sorted array and undersamples so that each class has an even number of samples
def undersample_from_sorted(sorted_list):
    # get number of samples in each class
    classes = len(sorted_list)
    count=[]
    # make empty arrays for counts in each class and counts for porting
    for i in range(classes):
        count.append(0.0)
    # get number of samples in each class
    for i in range(classes):
        count[i] = sorted_list[i].shape[0]
    # determine shape of new array
    features = sorted_list[0].shape[1]
    size = min(count)
    smallest = count.index(min(count))
    undersampled = np.empty(shape=(size*classes,features))
    # loop through each class
    index_count = 0
    for i in range(classes):
        # port smallest class directly
        if i==smallest:
            for j in range(size):
                undersampled[index_count,:] = sorted_list[i][j,:]
                index_count+=1
        # random sample the other classes
        else:
            samples = sorted_list[i].shape[0]
            for j in range(size):
                # get random sample from desired class
                current = np.random.randint(0,samples-1)
                undersampled[index_count,:] = sorted_list[i][current,:]
                # count up
                index_count+=1
    return undersampled

\end{lstlisting}

\section*{Declaration of Generative AI and AI-assisted technologies in the writing process}
During the preparation of this work the author(s) used ChatGPT in order to language editing and refinement. After using this tool/service, the author(s) reviewed and edited the content as needed and take(s) full responsibility for the content of the publication. [\href{https://www.elsevier.com/about/policies/publishing-ethics/the-use-of-ai-and-ai-assisted-writing-technologies-in-scientific-writing}{Elsevier Publishing Ethics}]

\section*{Acknowledgement}
The computational part of this work was supported by the U.S. Nuclear Regulatory Commission (NRC) and in part by the National Science Foundation (NSF) under Grant No. OAC-1919789.

\bibliographystyle{unsrtnat}
 \bibliography{references}

\begin{thebibliography}{33}
\providecommand{\natexlab}[1]{#1}
\providecommand{\url}[1]{\texttt{#1}}
\expandafter\ifx\csname urlstyle\endcsname\relax
  \providecommand{\doi}[1]{doi: #1}\else
  \providecommand{\doi}{doi: \begingroup \urlstyle{rm}\Url}\fi

\bibitem[Samal et~al.(2020)]{samal2020characterization}
Kumar Samal et~al.
\newblock Characterization and prediction of flow-conditions in the hot-leg of
  {PWR} during loss of coolant accident.
\newblock \emph{Nuclear Engineering and Design}, 359:\penalty0 110446, 2020.

\bibitem[Massaoudi et~al.(2021)]{massaoudi2021convergence}
Mohamed Massaoudi et~al.
\newblock Convergence of photovoltaic power forecasting and deep learning:
  State-of-art review.
\newblock \emph{IEEE Access}, 2021.

\bibitem[Demuth et~al.(2014)]{demuth2014neural}
Howard~B Demuth et~al.
\newblock \emph{Neural network design}.
\newblock Martin Hagan, 2014.

\bibitem[Lu et~al.(2021)]{lu2021deep}
Bi-Liang Lu et~al.
\newblock A deep adversarial learning prognostics model for remaining useful
  life prediction of rolling bearing.
\newblock \emph{IEEE Transactions on Artificial Intelligence}, 2\penalty0
  (4):\penalty0 329--340, 2021.

\bibitem[Leva et~al.(2017)]{leva2017analysis}
Sonia Leva et~al.
\newblock Analysis and validation of 24 hours ahead neural network forecasting
  of photovoltaic output power.
\newblock \emph{Mathematics and computers in simulation}, 131:\penalty0
  88--100, 2017.

\bibitem[Bin et~al.(2013)]{bin2013application}
Sun Bin et~al.
\newblock Application of gaussian process regression to prediction of thermal
  comfort index.
\newblock In \emph{2013 IEEE 11th International Conference on Electronic
  Measurement \& Instruments}, volume~2, pages 958--961. IEEE, 2013.

\bibitem[Yadav et~al.(2023)Yadav, Agarwal, Jain, Ramuhalli, Zhao, Ulmer,
  Carlson, Eskins, Nellis, Matrachisia, et~al.]{yadav2023state}
V~Yadav, V~Agarwal, P~Jain, P~Ramuhalli, X~Zhao, C~Ulmer, J~Carlson, D~Eskins,
  C~Nellis, J~Matrachisia, et~al.
\newblock State-of-technology and technical challenges in advanced sensors,
  instrumentation, and communication to support digital twin for nuclear energy
  application.
\newblock \emph{US Nuclear Regulatory Commission}, 2023.

\bibitem[Liu et~al.(2024)Liu, Alsafadi, Mui, O’Grady, and
  Hu]{liu2024development}
Yang Liu, Farah Alsafadi, Travis Mui, Daniel O’Grady, and Rui Hu.
\newblock Development of whole system digital twins for advanced reactors:
  Leveraging graph neural networks and sam simulations.
\newblock \emph{Nuclear Technology}, pages 1--18, 2024.

\bibitem[Kobayashi et~al.(2024{\natexlab{a}})]{kobayashi2024deep}
Kazuma Kobayashi et~al.
\newblock Deep neural operator-driven real-time inference to enable digital
  twin solutions for nuclear energy systems.
\newblock \emph{Scientific reports}, 14\penalty0 (1):\penalty0 2101,
  2024{\natexlab{a}}.

\bibitem[Kobayashi et~al.(2024{\natexlab{b}})]{kobayashi2024explainable}
Kazuma Kobayashi et~al.
\newblock Explainable, interpretable, and trustworthy ai for an intelligent
  digital twin: A case study on remaining useful life.
\newblock \emph{Engineering Applications of Artificial Intelligence},
  129:\penalty0 107620, 2024{\natexlab{b}}.

\bibitem[Kobayashi et~al.(2024{\natexlab{c}})]{kobayashi2024ai}
Kazuma Kobayashi et~al.
\newblock Ai-driven non-intrusive uncertainty quantification of advanced
  nuclear fuels for digital twin-enabling technology.
\newblock \emph{Progress in Nuclear Energy}, 172:\penalty0 105177,
  2024{\natexlab{c}}.

\bibitem[Kim et~al.(2014)]{kim2014prediction}
Dong~Yeong Kim et~al.
\newblock Prediction of leak flow rate using fuzzy neural networks in severe
  post-loca circumstances.
\newblock \emph{IEEE Transactions on Nuclear Science}, 61\penalty0
  (6):\penalty0 3644--3652, 2014.

\bibitem[Pan(2023)]{pan2023transfer}
Jie Pan.
\newblock Transfer learning for metal--organic frameworks.
\newblock \emph{Nature Computational Science}, pages 1--1, 2023.

\bibitem[Chen et~al.(2021)]{chen2021learning}
Chi Chen et~al.
\newblock Learning properties of ordered and disordered materials from
  multi-fidelity data.
\newblock \emph{Nature Computational Science}, 1\penalty0 (1):\penalty0 46--53,
  2021.

\bibitem[Kabir et~al.(2024)]{kabir2024transfer}
HM~Dipu Kabir et~al.
\newblock Transfer learning with spinally shared layers.
\newblock 163:\penalty0 111908, 2024.

\bibitem[Sola et~al.(1997)]{sola1997importance}
Jorge Sola et~al.
\newblock Importance of input data normalization for the application of neural
  networks to complex industrial problems.
\newblock \emph{IEEE Transactions on nuclear science}, 44\penalty0
  (3):\penalty0 1464--1468, 1997.

\bibitem[Do~Koo et~al.(2018)]{do2018prediction}
Young Do~Koo et~al.
\newblock Prediction of nuclear reactor vessel water level using deep neural
  networks.
\newblock In \emph{2018 International Conference on Electronics, Information,
  and Communication (ICEIC)}, pages 1--3. IEEE, 2018.

\bibitem[Radaideh et~al.(2020)]{radaideh2020neural}
Majdi~I Radaideh et~al.
\newblock Neural-based time series forecasting of loss of coolant accidents in
  nuclear power plants.
\newblock 160:\penalty0 113699, 2020.

\bibitem[El-Sefy et~al.(2021)]{el2021artificial}
M~El-Sefy et~al.
\newblock Artificial neural network for predicting nuclear power plant dynamic
  behaviors.
\newblock \emph{Nuclear Engineering and Technology}, 53\penalty0 (10):\penalty0
  3275--3285, 2021.

\bibitem[Griesemer et~al.(2023)]{griesemer2023accelerating}
Sean~D Griesemer et~al.
\newblock Accelerating the prediction of stable materials with machine
  learning.
\newblock \emph{Nature Computational Science}, pages 1--12, 2023.

\bibitem[Corrado(2021)]{corrado2021human}
Jonathan~K Corrado.
\newblock Human-machine system optimization in nuclear facility systems.
\newblock \emph{Nuclear Engineering and Technology}, 53\penalty0 (10):\penalty0
  3460--3463, 2021.

\bibitem[Kim et~al.(1993)]{kim1993application}
Wan~Joo Kim et~al.
\newblock Application of neural networks to signal prediction in nuclear power
  plant.
\newblock \emph{IEEE Transactions on Nuclear Science}, 40\penalty0
  (5):\penalty0 1337--1341, 1993.

\bibitem[Pickering et~al.(2022)]{pickering2022discovering}
Ethan Pickering et~al.
\newblock Discovering and forecasting extreme events via active learning in
  neural operators.
\newblock \emph{Nature Computational Science}, 2\penalty0 (12):\penalty0
  823--833, 2022.

\bibitem[Myung-Sub et~al.(1991)]{myung1991thermal}
Roh Myung-Sub et~al.
\newblock Thermal power prediction of nuclear power plant using neural network
  and parity space model.
\newblock \emph{IEEE transactions on nuclear science}, 38\penalty0
  (2):\penalty0 866--872, 1991.

\bibitem[Zhang et~al.(2019)]{zhang2019thermal}
Aoxin Zhang et~al.
\newblock Thermal power prediction of nuclear reactor core based on lstm.
\newblock In \emph{2019 Chinese Automation Congress (CAC)}, pages 5303--5307.
  IEEE, 2019.

\bibitem[Shouman et~al.(2022)]{shouman2022hybrid}
Marwa~A Shouman et~al.
\newblock A hybrid machine learning model for reliability evaluation of the
  reactor protection system.
\newblock \emph{Alexandria Engineering Journal}, 61\penalty0 (9):\penalty0
  6797--6809, 2022.

\bibitem[Kobayashi et~al.(2024{\natexlab{d}})]{kobayashi2024improved}
Kazuma Kobayashi et~al.
\newblock Improved generalization with deep neural operators for engineering
  systems: Path towards digital twin.
\newblock \emph{Engineering Applications of Artificial Intelligence},
  131:\penalty0 107844, 2024{\natexlab{d}}.

\bibitem[Lamarsh et~al.(2001)]{lamarsh2001introduction}
John~R Lamarsh et~al.
\newblock \emph{Introduction to nuclear engineering}, volume~3.
\newblock Prentice hall Upper Saddle River, NJ, 2001.

\bibitem[Kumar et~al.(2019)]{kumar2019influence}
Dinesh Kumar et~al.
\newblock Influence of nuclear data parameters on integral experiment
  assimilation using cook’s distance.
\newblock In \emph{EPJ Web of Conferences}, volume 211, page 07001. EDP
  Sciences, 2019.

\bibitem[Kumar et~al.(2022)]{kumar2022multi}
Dinesh Kumar et~al.
\newblock Multi-criteria decision making under uncertainties in composite
  materials selection and design.
\newblock \emph{Composite Structures}, 279:\penalty0 114680, 2022.

\bibitem[Dipu~Kabir et~al.(2010b{\natexlab{a}})]{kabir2010theory}
Hussain~Mohammed Dipu~Kabir et~al.
\newblock \emph{A Theory of Loss-less Compression of High Quality Speech
  Signals with Comparison}, pages 136--141.
\newblock 2010b{\natexlab{a}}.

\bibitem[Dipu~Kabir et~al.(2010a)]{kabir2010non}
Hussain~Mohammed Dipu~Kabir et~al.
\newblock \emph{Non-linear down-sampling and signal reconstruction, without
  folding}, pages 142--146.
\newblock 2010a.

\bibitem[Dipu~Kabir et~al.(2010b{\natexlab{b}})]{kabir2010watermarking}
Hussain~Mohammed Dipu~Kabir et~al.
\newblock \emph{Watermarking with fast and highly secured encryption for
  real-time speech signals}, pages 446--451.
\newblock 2010b{\natexlab{b}}.

\end{thebibliography}

\end{document}